\documentclass[10pt,journal,compsoc]{IEEEtran}
\usepackage{tabularx}
\usepackage{multirow} 
\usepackage{threeparttable}
\usepackage{float}
\usepackage{graphicx}
\usepackage{subfigure}
\usepackage{amssymb}
\usepackage{times}
\usepackage{soul}
\usepackage{url}
\usepackage[hidelinks]{hyperref}
\usepackage[utf8]{inputenc}
\usepackage[small]{caption}
\usepackage{enumitem}
\usepackage{graphicx}
\usepackage{amsmath}
\usepackage{amsthm}
\usepackage{booktabs}
\usepackage{algorithm}
\usepackage{algorithmic}
\ifCLASSOPTIONcompsoc
  \usepackage[nocompress]{cite}
\else
  \usepackage{cite}
\fi
\ifCLASSINFOpdf
\else
\fi
\begin{document}
%
\title{Feature Noise Resilient for QoS Prediction with Probabilistic Deep Supervision}
%
%
%
%

\author{Ziliang~Wang, Xiaohong~Zhang, Ze Shi Li, Sheng Huang, Member, IEEE and  Meng~Yan, Member, IEEE

\IEEEcompsocitemizethanks{


\IEEEcompsocthanksitem Ziliang Wang is with Key Laboratory of High Confidence Software Technologies (Peking University), Ministry of Education; School of Computer Science, Peking University, Beijing, China.

Ze Shi Li is in the Department of Computer Science at the University of Victoria, Canada

Xiaohong~Zhang, Meng~Yan are with Key Laboratory of Dependable Service Computing in Cyber Physical Society (Chongqing University),  Ministry of Education, China and School of Big Data and Software Engineering, Chongqing University, Chongqing 401331, China. \protect

\IEEEcompsocthanksitem 
Xiaohong Zhang is the corresponding authors.\\
E-mail: xhongz@cqu.edu.cn
}
}
%



\IEEEtitleabstractindextext{%
\begin{abstract}
Accurate Quality of Service (QoS) prediction is essential for enhancing user satisfaction in web recommendation systems, yet existing prediction models often overlook feature noise, focusing predominantly on label noise. In this paper, we present the Probabilistic Deep Supervision Network (PDS-Net), a robust framework designed to effectively identify and mitigate feature noise, thereby improving QoS prediction accuracy. PDS-Net operates with a dual-branch architecture: the main branch utilizes a decoder network to learn a Gaussian-based prior distribution from known features, while the second branch derives a posterior distribution based on true labels.
A key innovation of PDS-Net is its condition-based noise recognition loss function, which enables precise identification of noisy features in objects (users or services). Once noisy features are identified, PDS-Net refines the feature’s prior distribution, aligning it with the posterior distribution, and propagates this adjusted distribution to intermediate layers, effectively reducing noise interference.
Extensive experiments conducted on two real-world QoS datasets demonstrate that PDS-Net consistently outperforms existing models, achieving an average improvement of 8.91\% in MAE on Dataset D1 and 8.32\% on Dataset D2 compared to the ate-of-the-art. These results highlight PDS-Net’s ability to accurately capture complex user-service relationships and handle feature noise, underscoring its robustness and versatility across diverse QoS prediction environments.
\end{abstract}

}

\maketitle

\IEEEdisplaynontitleabstractindextext

%
\IEEEpeerreviewmaketitle

\IEEEraisesectionheading{\section{Introduction}\label{sec:introduction}}
\IEEEPARstart{W}{ith} the development of contemporary cloud computing and distributed technology, Web services are becoming an indispensable part of service-oriented industrial application architecture.
Due to the characteristics of cross-region and distributed service deployment, there are complex situations between different users and services.
At the core of this effort is the capability to recommend high-quality services~\cite{mouli2016web}.
Beyond their functional capabilities, the non-functional attributes, which are symbolize of a Web service's quality, play a pivotal role in underpinning the reliability of the Web services. 
These attributes, collectively denominated as the Quality of Service (QoS), encapsulate various metrics such as response latency, invocation failure rate, throughput, capacity and robustness~\cite{menasce2002qos}.
For example, the concept of response time based service recommendation involves forecasting the response times for services  that are not yet used and prioritizing the services with the best response.
However, recommending services that do not meet the QoS standards may lead to longer response times and thus potentially lead to a very poor user experience.
This underscores the importance of accurately predicting missing QoS metrics using current observations to ensure effective service recommendations~\cite{HeNCF,huang2016deep}.

Quality of Service (QoS) prediction approaches in existing literature are primarily categorized into two types: Collaborative Filtering (CF)-based approaches~\cite{carlkadie1998empirical,13,18,26} and Deep Learning (DL)-based models~\cite{86,21,zou2022ncrl}. 
The former typically relies solely on the quality similarity of users or services and exhibits lower computational complexity. 
But they are difficult to use the contextual features, consequently often falling short in achieving sufficient prediction accuracy.
Conversely, DL-based prediction methods extract nonlinear feature information from service and user features, such as user location and network information.
Although these methods have higher training cost, but can make full use of the known features to achieve high-precision QoS prediction.

To enhance the use of features and improve prediction accuracy further, recent studies have introduced prediction models employing deep network architectures, including structures like ResNet~\cite{zou2022ncrl,zhang2021probability}.
For example, Zou et al. proposed a two-tower deep residual network to learn the deep features of users and services\cite{zou2022ncrl}. 
%
In this model, known features of users and services are superimposed and propagated through the network's intermediate layers via residual blocks.

However, using deeper network structures has its benefits, but the challenge we encounter when it comes to real-world applications \textbf{the feature noise also expands with the deep network structure}~\cite{xie2019feature}.
Recent literature has highlighted the challenge of "noisy" during QoS prediction, which hinders prediction performance~\cite{ye2021outlier,lu2023feature}. 
These research mainly focuses on alleviating the impact of noisy in the label on the training of prediction models~\cite{ye2021outlier,miliauskaite2023effect}.
As an example, Ye et al. proposed to use the isolation forest algorithm to remove points in labels that may be outliers(as noise)~\cite{ye2021outlier}.
However, the noise in the features has not been given sufficient attention.
Feature noise can interfere with the learning of the true patterns in the data, making it difficult for the model to identify useful features and the relationships between QoS and features. 
In the field of QoS prediction, feature noise has been a long-standing issue, which hinders effective improvements in prediction accuracy.
These feature noise data may be come from inaccuracies in data collection or from malicious activities during service access, such as deceptive information or manipulated gateway usage~\cite{mouli2016web}.
During the exploration in real web data~\cite{zheng2008ws}, two main types of noisy features were discovered:
\begin{enumerate}
\item Incorrect features: Incorrect features predominantly arises from errors in data collection methods or intentional falsification by the subjects under investigation~\cite{mouli2016web}.
Additionally, the variability introduced by dynamic IP systems can alter users' network information. Notably, the acquisition error rate of RT in the WS-Dream dataset reaches 5.11\%~\cite{3}.
\item Missing features: Missing features often results from data acquisition failures or privacy measures that prevent the collection of critical data~\cite{zhang2023deep}. In the WsDream open-source dataset, such instances are denoted as "NULL"~\cite{zheng2008ws}. Taking the real service dataset WS-dream as an example, 1214 services out of 5825 collected services have missing features.
\end{enumerate}

We introduce PDS-Net, a novel QoS prediction approach leveraging Gaussian-based probabilistic networks, which enables enhanced supervision across intermediate layers to mitigate noise interference effectively.
Our approach is based on the concept that a Gaussian-based probabilistic network can provide superior supervision for intermediate layers, thereby mitigating the impact of noise features in the QoS prediction task. 
The architecture of PDS-Net incorporates a Gaussian-based probabilistic neural network that first obtains a prior distribution using the known features distilled by a probabilistic decoder.
Then, deep uncertain features that sample from prior distribution are subsequently amalgamated with known features to yield QoS predictions. 
Moreover, the PDS-Net constructs a posterior distribution grounded in true labels via supervised  probabilistic decoder.
The key of the method is the conditional supervised learning of the prior distribution and the posterior distribution.
We propose a condition-based noise recognition loss function to judgment.
To be specific, when the prediction error based on the prior feature is too large, we have reason to believe that there is noise in this feature.
To this end, by aligning the prior distribution with the posterior distribution that form true labels, we realize the elimination of the noise feature distribution.
The alignment is performed by minimizing the Kullback-Leibler distance between two distributions.

Our contribution to DL-based QoS prediction through the probabilistic supervision network is twofold. 
Firstly, we propose probabilistic supervision networks to supervise QoS features and attenuate the adverse effects of noise from data collection differences or malicious data intrusion.
Secondly, our method can also effectively provide referable latent feature distribution for users and services that lack features.
Typically, this missing features is bridged with null values or zeroes. During training, PDS-Net employs posterior distributions premised on genuine labels to guide prior distributions extracted from the missing data, eliminating errors attributed to data deficiency and offering invaluable deep latent features for such instances.

In summary, the main contributions of this paper are as follows:
\begin{itemize}
\item [a)] We propose the PDS-Net, a novel probabilistic supervision network. By employing a Gaussian-based probabilistic paradigm, this network enhances the resilience of deep supervision architectures against feature noisy training data, achieving this without any additional computational overhead during inference.
\item [b)]  We introduce a seamless integration of probabilistic networks with supervised learning. This approach includes a novel conditional loss function to distinguish noisy data, mitigate the impact of noise feature.
\item [c)] We exhaustively evaluate PDS-Net against existing benchmarks on two real-world datasets. The experiment proved that accentuates PDS-Net's superiority in QoS prediction precision.\footnote{The code and experimental record open source address will be updated upon acceptance of the paper}
\end{itemize}

\section{Related work}\label{sec:Related work}
This section sheds light on probabilistic neural networks and deep supervised networks, providing an introduction to their current applications.

\textbf{Probabilistic Neural Network.} Probabilistic neural networks (PNNs) offer a viable alternative to classical back-propagation networks, exempting them from exhaustive forward and backward computations. 
Existing studies have shown that probabilistic networks perform well even with limited data\cite{mohebali2020probabilistic}.
The first goal of pnn is to produce a singular continuous value;
they aspire to determine distribution parameters, like a Gaussian distribution's mean and variance, predicated on certain input features.
The utilization of distribution-based prediction methods offers a plethora of advantages compared to other approaches, with one notable benefit being the provision of uncertainty bounds for predictions.
A paramount uncertainty source in QoS data is observational noise, which can manifest as spurious location data or inaccurate network information. 
Analyzing these data distributions facilitates the assessment of such uncertainties. 
For instance, the Social LSTM model predicts a two-dimensional location Gaussian distribution's parameters, encapsulating probabilistic user behavior modeling~\cite{alahi2016social}. 
To adeptly discern the posterior distribution utilizing a Gaussian paradigm, Kingma and Welling advanced a stochastic variational inference algorithm termed VAE (Auto-Encoding Variational Bayes)~\cite{kingma2013auto}.

\textbf{Deep Supervision Network.}
The optimization of deep neural networks can be challenging due to the large number of intermediate layers.  
In the existing QoS prediction research, Resnet is a common method to alleviate the problem of vanishing gradients in the middle layer.
Similarly, deep supervised net was proposed by Lee et al. to supervise these layers \cite{lee2015deeply}. 
Wang et al. showed that deep supervision can improve performance and mitigate the problem of vanishing gradient \cite{wang2015training}. 
Recently, studies have proposed using knowledge distillation to minimize the difference between the final classifier and intermediate classifiers, such as dynamic distillation \cite{li2020dynamic}.
Zhang et al. have proposed a supervisory network that uses augmentation-based contrastive learning to supervise intermediate layers \cite{zhang2022contrastive}. 
Previous research has demonstrated the effectiveness of deep supervision methods in tasks such as semantic segmentation \cite{zhang2018deep}, dynamic neural networks \cite{zhang2019scan}, knowledge distillation \cite{zhang2020task}, and object detection \cite{lee2015deeply}.

Prevalent QoS prediction stratagems bifurcate into two realms: collaborative filtering-based methods\cite{hussain2022new,muslim2022s,wu2020data,chowdhury2020cahphf,liu2019context,zheng2020web} and those rooted in deep learning\cite{liang2021recurrent,li2021topology,xia2021joint,ghafouri2020survey}. 

\textbf{Collaborative Filtering-Based QoS prediction Approaches.} 
Collaborative Filtering (CF) is a prevalent technique for QoS prediction that utilizes similar users or services to predict the QoS of a target request~\cite{1,2,3,4}. CF-based approaches can be classified into two categories: memory-based and model-based. The former utilizes the similarity between users or services to estimate the QoS, while the latter applies machine learning models to capture the relationship between users or services and QoS.

Memory-based CF approaches predominantly utilize quality of service metrics, like response time and network traffic, as well as user and service attributes, including location information, to ascertain similarity between users or services. The foundational principle entails calculating similarity scores relative to the target object. Such similarities might be computed using user-centric techniques, e.g., UPCC~\cite{12}, service-based methods like IPCC~\cite{13}, or amalgamations of user and service attributes as seen in UIPCC~\cite{14}. These strategies, although simplistic, are efficient since they harness a singular attribute to determine the similarity between users and services.
To further enhance prediction accuracy, Bellcore and colleagues incorporated contextual details, leveraging evaluations from analogous users to forecast the target user's service evaluation\cite{79}.
Beyond memory-centric strategies, model-based techniques, such as matrix decomposition (MF)\cite{8,9}, have found prevalent use in QoS prediction. To bolster the precision of CF, contemporary model-based methods emphasize the integration of context, like time and location. For instance, Zhang et al. introduced a time-aware framework that personalizes QoS value prediction in sync with the service user's specific timeline\cite{35}. Alternatively, Wang and colleagues suggested a distance-oriented selection tactic, harnessing user coordinates~\cite{81}, whereas Chen et al. devised RegionKNN models, clustering users based on IP addresses and location similarity~\cite{2}. Nevertheless, concerns surrounding data collection costs and user privacy constrain the obtainability of such contextual details. To this end, latent factor (LF)-oriented QoS predictors, epitomized by Luo et al.'s nonnegativity constraint-based latent factor (NLF) model~\cite{29}, are increasingly favored for their scalability and precision.

\textbf{Deep Learning-Based QoS prediction  Approaches.}
In order to enhance the nonlinear learning capability of collaborative filtering, He et al. pioneered a paradigm named Neural network-based Collaborative Filtering (NCF), which supplanted the inner product with a neural construct skilled in distilling arbitrary functions from data\cite{20}. Additionally, several deep learning-centric prediction techniques have been formulated to refine QoS prediction accuracy.

Furthermore, the incorporation of spatio-temporal data has been observed in QoS prediction efforts. A case in point is the model by Zhou et al., which equips each temporal slice with a latent attribute to elucidate its state~\cite{83}. Bayesian-based probabilistic networks have been embraced for QoS prediction assignments as well. For instance, Wang et al. fashioned a motif-centric dynamic Bayesian network, mastering conditional dependencies throughout time intervals for QoS value prognostication \cite{wang2016online}. While intricate deep models do elevate accuracy, they necessitate abundant features for predictive conditions. Xiong and colleagues championed the Deep Hybrid Service Recommendation (DHSR) approach, harnessing text congruence and an MLP network to decipher the nonlinear ties between services and mashups \cite{xiong2018deep}. Moreover, to unearth latent data, Wang and team propounded a latent state model, exploiting latent factor algorithms to extract diverse user and service latent attributes, thereby enhancing deep neural network model prediction precision \cite{86}. It's pertinent to note that QoS prediction methods grounded in deep learning hinge profoundly on potent data attributes to forge high-dimensional affiliations between features and QoS. Analogous to collaborative filtering techniques, data noise can impair model precision.

PDS-Net gleans insights from other realms, including models predicated on the Denoising Autoencoder (DAE)\cite{im2017denoising} and the Contrastive Deep Supervision\cite{zhang2022contrastive} techniques. DAEs aim to reconstruct inputs post noise-infusion at input stages, ameliorating the repercussions of noise. 
Concurrently, the Contrastive Deep Supervision methodology, as suggested by Zhang et al., provides additional supervision to the intermediate layer via contrastive learning, enhancing overall network efficiency. 
A salient differentiation between PDS-Net and these models is PDS-Net's abstention from deliberate noise introduction to inputs. 
Contrarily, it supervises the intermediate stratum employing the probability distribution intrinsic to the genuine labels, thereby mitigating errors stemming from noisy attributes.

\section{Approach}
\begin{figure*}
\centering
\includegraphics[width=0.98\textwidth]{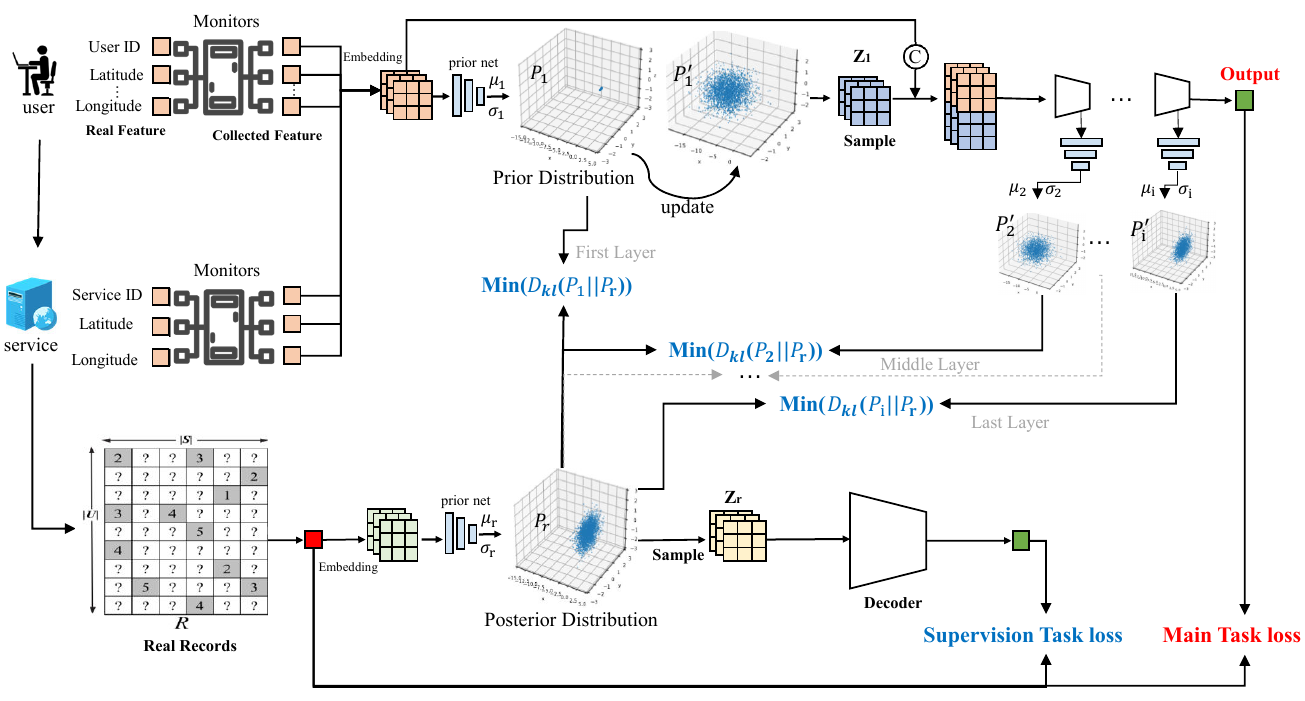}
\captionsetup{justification=centering}
\caption{The framework of PDS-Net. Prior net is a decoder network; $P_1$ is a Gaussian-based prior distribution learned from known information; $P_2$ and $P_i$ are priori distributions based on deep variable in the middle layer. The posterior distribution $P_r$ is learned from real label.}
\label{fig:pds}
\end{figure*}

As depicted in \autoref{fig:pds}, the network architecture we propose consists of two main branches: one is a backbone network configured as a probabilistic neural network utilizing known features, while the second branch is similarly structured but employs real labels. 
These branches are integrated through a conditional loss function that implements the feature noise resistance process.
\subsection{Feature Embedding}
The step objective is to augment neural networks' capacity to assimilate additional data features. 
We integrate data features(identifiers for users and services, location data) and true labels (QoS values) into a Keras embedding layer. 
The embedding layer acts as a bias-free fully connected layer. It one-hot encodes inputs into zero vectors of a specific dimension, only activating the i-th position of said vector. 
The embedding strategy we adopt mirrors the dense vector representations common in natural language processing, mapping categorical data onto dense embedding vectors in high-dimensional space. 
This mapping process is outlined below:
\begin{equation}
\centering
U^{k}=f(P*U_{i}+b_i).
\end{equation}
\begin{equation}
\centering
S^{k}=f(L*S_{i}+b_i).
\end{equation}
\begin{equation}
\centering
A^{k}_{u}(A^{k}_{s})=f(T_{i}A_{u_i}(A_{s_i})+b_i).
\end{equation}
\begin{equation}
\centering
C^{k}_{u}(C^{k}_{s})=f(T_{i}C_{u_i}(C_{s_i})+b_i).
\end{equation}
 $U^{k}_{i}$ and $S^{k}_{i}$ represent k-dimensional embedding vectors for users and services, respectively. 
 $A^{k}_{u}$ and $A^{k}_{s}$ are the k-dimensional embedding vectors for the user's and service's AS data.
 $C^{k}_{u}$ and $C^{k}_{s}$ are the k-dimensional embedding vectors for the city data of users and services.
$P$ and  $L$  are weight matrices for user and service id embeddings.
$L_i$ is the weight matrix for location features embeddings.
$b_i$  is the bias term initialized to zero.
$f$ is the activation function for this layer.

Subsequent to this, we amalgamate the individual feature vectors to create a composite feature vector, given by:
\begin{equation}
C=\Phi (U^{k},S^{k},A^{k}_{s},A^{k}_{u},C^{k}_{s},C^{k}_{u}).
\end{equation}
where $\Phi$ represents the concatenation operation in Tensorflow, executed row-wise (axis=1).

\subsection{Forecasting with Prior Distribution}
%
The backbone network implements QoS prediction through a deep probabilistic neural network.
PDS-Net constructs a Gaussian prior distribution from middle layer vector and generates uncertainty features through random sampling. 
The main component of our architecture is the prior Gaussian prior distribution $\mathbb{R}^{N}$, where each position encodes a different variant (e.g., N = 64, which yielded the best performance in our experiments).
The 'prior net', parameterized by weights $w$, estimates the probability of these variants for a given combined feature $C$. 
The detailed structure of the 'prior net' is as follows:
\begin{equation}\label{pri}
E_i=Linear(f^{[N*2]}(Relu(f^{[k]}(C))).
\end{equation}
where $f^{[k]}$ signifies a fully connected layer with $k$ neurons.
$Relu$ function is a ReLU activation function that zeroes out negative values.
$Linear$ function is a linear activation that all value remains unaltered.

From specific variants  $E_i$, we calculate the mean of the multivariate normal distribution $\mu _i$ and  the diagonal values of the covariance matrix for the multivariate normal distribution $\sigma_i$:
\begin{eqnarray}
\small
\mu_i &=f_{\mu}(E_i; \mathbf{W}_{\mu}), \\
\sigma_i &=f_{\sigma}(E_i; \mathbf{W}_{\sigma}).
\end{eqnarray}
Then we obtain the Gaussian distribution $P_i$ by MultivariateNormal function from tensorflow\_probability function: 
\begin{equation}
 P_i(\cdot |C)=\mathcal{N}(\mu _i,diag(\sigma_i)).
\end{equation}
where diag function takes one-dimensional vectors and forms a matrix with one-dimensional arrays as diagonal elements.  
This function takes the expectation and covariance matrix of a Gaussian distribution to construct that Gaussian distribution.

\noindent\textbf{Disentangled Latent Variable Sample From Prior Distribution.} 
PDS-Net obtains uncertainty deep features from the $P_i$ by random sampling.
\begin{equation}
 Z_{i}| P_{i}^{'} (\cdot |C)\sim Gaussian(C).i=1,2,3
\end{equation}
The sampling is executed using TensorFlow's random sampling function. The shape of  $Z_{i}$ aligns with that of $\mu$ and $\sigma$. Subsequently, the disentangled latent variable and know features merge to form a new fused feature.
\begin{equation}
 Z=\Phi(Z_{i},C).
\end{equation}

In its final stages, PDS-Net processes these through multiple shared, fully-connected layers to yield intermediate results 
 $Z$, which aids in understanding the nonlinear interrelations among features:
\begin{equation}
x_i=f^{[2^{v}]}(Z; \mathbf{W}_x),i=1,2,3,v=10,9.
\end{equation}
\begin{equation}
\hat{y_1}=f^{1}(x_i; \mathbf{W}_x).
\end{equation}
In the above,  $\Phi$ symbolizes the merge operation, while  $F$  stands for the flattening process executed by Keras' Flatten function. 
$f^{[2^{v}]}$ indicates a fully connected layer, $\hat{y_1}$ is the predicted value derived from the known information and the disentangled latent variable.

\subsection{Learning with the Posterior Distribution}
In this section, we present how to train the posterior distribution $P_r$ from the true labels.
The detailed network structure is provided below:
\begin{equation}
E_r=Linear(f^{[N*2]}(Rule(f^{[512]}(y))).
\end{equation}
\begin{eqnarray}
\small
\mu_r &=f_{\mu}(E_r; \mathbf{W}_{\mu}), \\
\sigma_r &=f_{\sigma}(E_r; \mathbf{W}_{\sigma}).
\end{eqnarray}
where $y$ represents the actual label. Ultimately, the posterior distributions $Pr$ are derived through the 'prior net'.
\begin{equation}
P_r(\cdot |y)=\mathcal{N}(\mu _r,diag(\sigma_r)).
\end{equation}

\noindent\textbf{Disentangled Latent Variable Sample From Posterior Distribution.} 
For this part, PDS-Net navigates through several shared, fully-connected layers to obtain the intermediate $Z_r$ results, focusing on unveiling the nonlinear relations between features:
\begin{equation}
 Z_{r}| Pr(\cdot |y)\sim Gaussian(y).
\end{equation}
\begin{equation}
x_{m_i}=f^{[2^{v}]}(Z; \mathbf{W}_x).
\end{equation}
\begin{equation}
\hat{y_2}=f^{1}(x_{m_i}; \mathbf{W}_x).
\end{equation}
Where $\hat{y_2}$ represents the prediction value grounded on the real label.

\subsection{Noise-Resilient Process}
Based on the conditional loss function proposed in Sec 3.5, the model selectively triggers the Noise-Resilient Process.
In the backbone network, we have established prior distributions $P_i$ for deep uncertainty features. Meanwhile, the posterior distributions $P_r$ is established in the supervised network respectively. 
In the noisy-resilient process,  by minimizing the Kullback-Leibler divergence between $P_r$ and $P_i$, the prior distribution $P_i$ based on known features is made closer to the posterior distribution.  
\begin{equation}
P_{i}^{'} = f_{Min}(D_{kl}(Pr||P_i)))=f_{Min}(\mathbb{E}_{z\sim Pr_i}[logPr-logP_i])).
\end{equation}

\subsection{Loss Function and Training Strategy}
\textbf{Task Loss.}  In the realm of regression tasks, various loss functions abound. While some cater to classification, others are tailored for regression. 
Given the nature of the QoS prediction problem—a regression problem—standard loss functions like the Mean Square Error (MSE), Mean Absolute Error (MAE), and Huber functions are evaluated. 
Through extensive comparative experimentation, the MAE function emerged as superior. 
It is mathematically captured as:
\begin{equation}
\centering
T_i(y,\hat{y_x})=\frac{1}{n} \sum_{i=1}^{n}\left |y_{i}-\hat{y_x}_{i} \right |_{abs}.
\end{equation}
\textbf{Hierarchical KL Loss.} Beyond the prediction task loss, the model is also sensitive to the Kullback-Leibler divergence. 
This divergence signifies the penalty arising from deviations between the prior distribution 
 $P_i$ and the posterior distribution $Pr$.  
 It is defined as:
\begin{equation}
\begin{aligned}
D_{KL_i}(P_i,P_r)=\mathbb{E}_{z\sim P_{r_i}}[logP_r-logP_i].
\end{aligned}
\end{equation}
\textbf{Noise Feature Discrimination Mechanism and Training Strategy.} 
The initial step in the PDS-Net methodology discerns the presence of noise in data. First, the backbone network processes known features and labels. 
If no significant prediction difference is found by the training scheme, it indicates trust in the data features. In such circumstances, PDS-Net strategically avoids refining the backbone network with probabilistic deep supervision network.
On the contrary, if the prediction of QoS values by known features produces a large error, the known feature distribution $P_i$ is optimized by a probabilistic deep supervised structure. 
It aims to learn the probability space, learned from authentic labels, within the feature domain, thereby retroactively inferring the genuine feature space. 

To implement the above training strategy, the final loss function is defined as follows.
\begin{equation}                                                        
H_{loss}=\left\{\begin{matrix}
T_(\hat{y_1}) +T_(\hat{y_2}) &  |y-\hat{y_i}|_{abs}<\delta .\\ 
 \lambda_1 D_{KL_1}+\lambda_2 D_{KL_2}...+\lambda_i D_{KL_i} &  else .
\end{matrix}\right.
\end{equation}
$\delta$ is the core hyperparameter, which is used to judge whether the features of the data are trustworthy.
If the prediction error is greater than $\delta$, we think that it is difficult to map from the known features to the real label, and the data features may be contaminated.
The training process starts with a initialized weights(Sec 5.6.1), and during training, the $D_{KL_i}$ encourages the prior distribution $P_i$ and the posterior distribution $P_r$ to approach each other.
During testing, a random array (e.g., an array of all ones) is used instead of $\hat{y}$ to prevent the influence of $P_r$ on the test results, as the network parameters are not updated during testing.
$\lambda_i$ are hyperparameter, which are manually tuned according to different datasets. 
Depending on the prediction task and the depth of the model architecture, different levels of deep supervision will have different effects.
It is worth noting that after ablation experiments, the maximum value of $i$ is set to 3 in this paper, that is, through 3 times of deep supervision.

\section{STUDY SETUP}\label{sec:STUDY SETUP}
This section provides a detailed introduction to our experimental setup, including information on the datasets used and the evaluation metrics employed. All our experiments were conducted using Python 3.7.3 and Tensorflow 2.4.0. The deep network was implemented using Tensorflow\_probability 1.3.0, while the baselines were implemented using sklearn. We conducted all experiments on a server equipped with two 2.4GHz Intel Xeon CPUs and 256GB of memory running Ubuntu 18.04. We used a batch size of 64 and the Nadam optimizer for training.

\subsection{Datasets}
We evaluate our method on two benchmark datasets, namely the Web service QoS data collected by the WS-Dream system\cite{3}. 
We partitioned the dataset into multiple granularities to better simulate the reality of data sparsity, following the approach used in previous studies.
The segmentation method used is illustrated in Figure 1. The datasets are described as follows:

\textbf{Dataset D1.} This is a real-world Web services dataset collected by Zheng et al.~\cite{3}, which contains 1,974,675 QoS values of Web services from 339 users on 5,825 services. It also includes the location information of users and services. 
In this paper, we represent this dataset as a user-service matrix, where each row represents a user, each column represents a service, and each value in the matrix represents the corresponding response time (RT).

\begin{figure} 
\centering 
\subfigtopskip=2pt 
\subfigbottomskip=2pt 
\subfigure[Distribution of Dataset D1]{\includegraphics[width=0.49\linewidth]{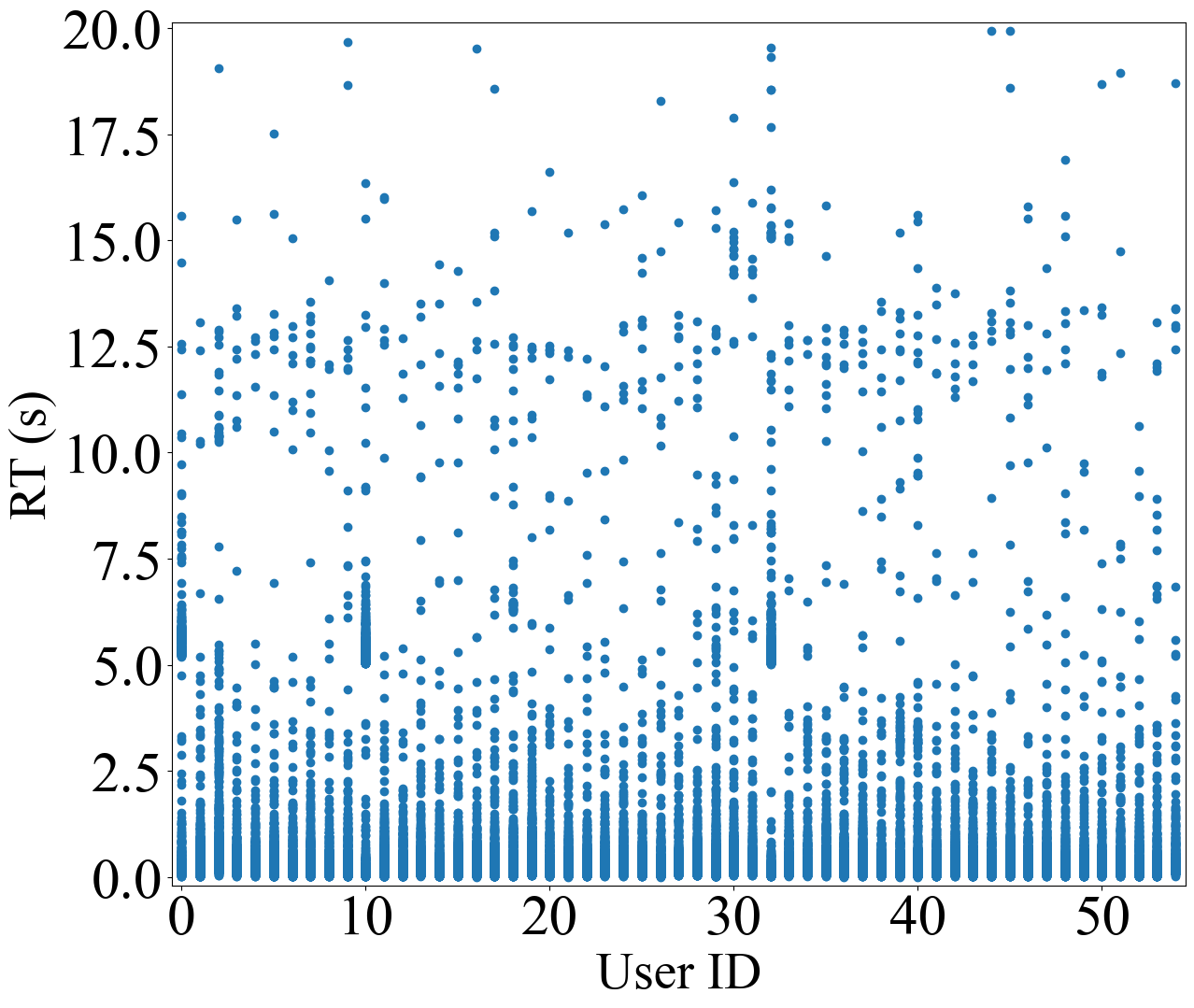}}
\subfigure[Distribution of Dataset D2]{\includegraphics[width=0.49\linewidth]{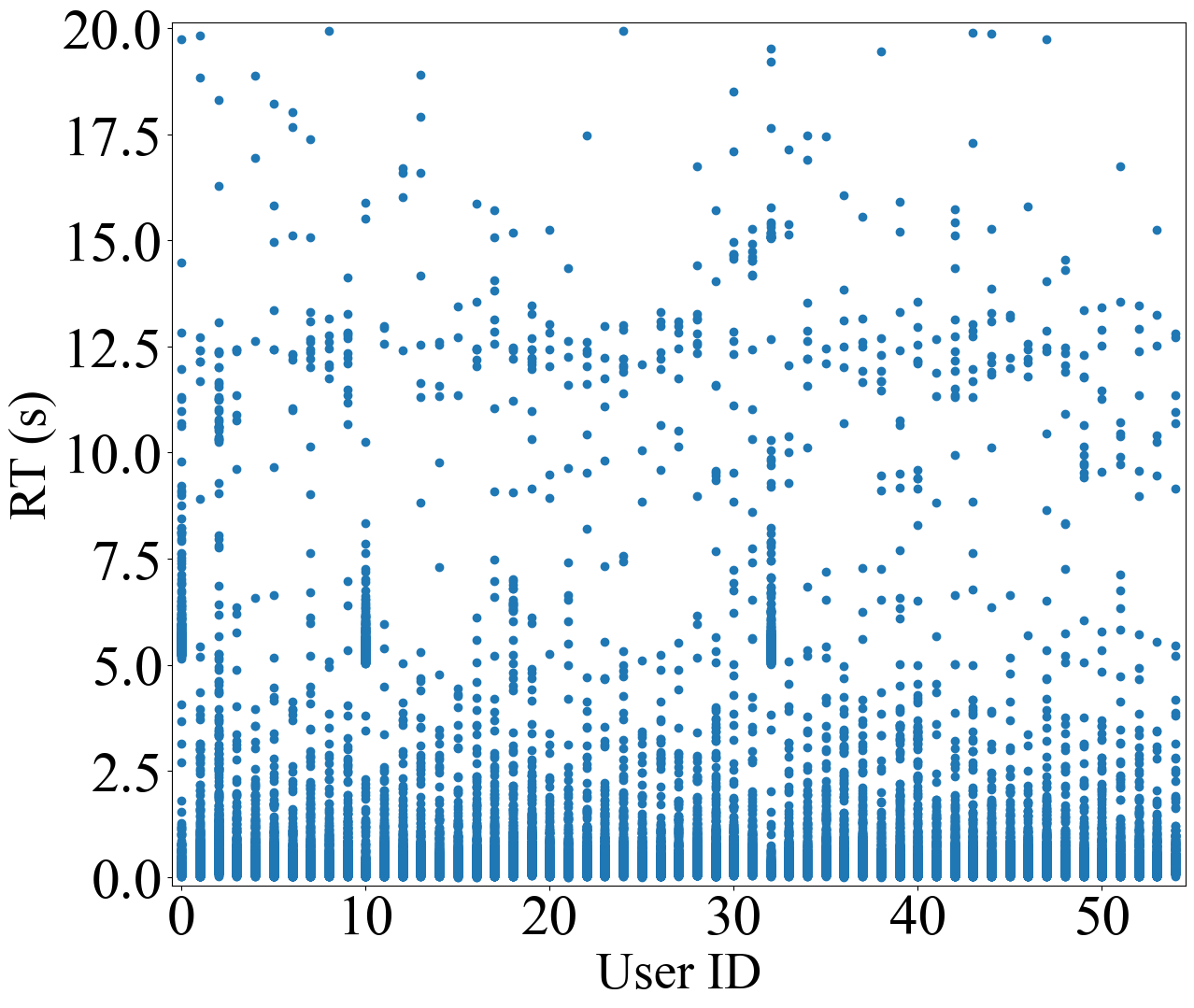}}
\vspace{-1em}
\caption{QoS distribution of datasets}
\vspace{-1em}
\label{fig:data}
\vspace{-1em}
\end{figure}

\textbf{Dataset D2.}  This dataset is obtained from a recent study~\cite{77}, where outliers were removed from the original WS-Dream dataset using the iForest (isolation forest) method for outlier detection. 
The iForest method calculates an outlier score for each datum, which takes a value in the range of [0, 1], with a larger value indicating a higher possibility of being an outlier. 
In this study, we set the outlier score threshold to 0.1, following~\cite{77}. The difference between this dataset and Dataset D1 is shown in \autoref{fig:data}.

\textbf{Dataset D1 with added noise.} We generated this dataset based on Dataset D1. As shown in Table 2, missing information is easy to detect, but currently, there is no effective method to distinguish incorrect  information.
We show in Figure 3 the presence of noise in the existing data, because of the lack of feature labels, and we show the part of missing features.
To further investigate the effect of noise on the prediction model's performance, we introduced noise artificially. Specifically, we randomly selected 10\% of the users and provided them with fake city and Autonomous System (AS) information.

\begin{table}
\centering
\caption{Properties of all the designed test cases.}
\vspace{-1em}
\begin{tabularx}{9cm}{lllXXX}
\hline
\hline
No.&Density&Train:Test:Validation&Train&Test&Validation\\
\hline
D1.1&0.05 & 5\%:75\%:20\% & 98,721 &1,399,535&374,564 \\
D1.2&0.10 & 10\%:70\%:20\% & 197,440 &1,301,602&374,564\\
D1.3&0.15& 15\%:65\%:20\% & 296,182 &1,203,007 &374,564\\
D1.4&0.20 & 20\%:60\%:20\% & 394,926 &1,104,303& 374,564\\ 
\hline
D2.1&0.02 & 2\%:78\%:20\% & 37,375 &1,310,535 &369,638\\
D2.2&0.04 & 4\%:76\%:20\% & 74,969&1,572,292&369,638\\
D2.3&0.06& 6\%:74\% :20\%& 1,12,016  &1,206,517&369,638 \\
D2.4&0.08 & 8\%:72\% :20\%& 1,50,071 &1,172,461  &369,638\\
D2.5&0.10 & 10\%:70\% :20\%& 1,86,059  &1,140,269& 369,638\\
\hline
\end{tabularx}
\end{table}

\begin{table}
\centering
\begin{threeparttable}
\fontsize{6}{7}\selectfont
\caption{The example of Ws-Dream data}
\label{tab:performance_comparison}
\setlength{\tabcolsep}{1.4mm}{
\begin{tabular}{|c|c|c|c|c|c|c|}
\hline
\multicolumn{7}{|c|}{The Dateset of Ws-Dream}\cr
\hline
\hline
User ID & Service ID & RT&User-Country& User-AS & Service-Country & Service-AS\cr\hline
0&0&5.982& United States&AS7018 AT&United States&AS3356\cr
2&0&2.13& Japan&NTT C.C&United States&AS3356\cr
6&3&0.854&United States&AS131&United States&AS1728\cr
...&...&...& ...&...&...&...\cr
239&3945&0.132&Switzerland&AS559&United States&null\cr
...&...&...& ...&...&...&...\cr
339&5825&0.44&Slovenia&AS2107&Australia&null\cr
\hline
\end{tabular} }
\end{threeparttable}
 
\end{table}

\begin{table*}[tp]
\centering
\caption{Performance Comparison with Different Training Ratios on Dataset D1 (Best Results in Bold Numbers)}
\label{tab:performance_comparison}
\begin{threeparttable}
\setlength{\tabcolsep}{5.5mm}
\begin{tabular}{c c c c c c c c c}
\toprule
\textbf{Method} & \multicolumn{2}{c}{\textbf{D1.1}} & \multicolumn{2}{c}{\textbf{D1.2}} & \multicolumn{2}{c}{\textbf{D1.3}} & \multicolumn{2}{c}{\textbf{D1.4}} \\ 
\cmidrule(lr){2-3} \cmidrule(lr){4-5} \cmidrule(lr){6-7} \cmidrule(lr){8-9}
& MAE & RMSE & MAE & RMSE & MAE & RMSE & MAE & RMSE \\ 
\midrule

UPCC & 0.634 & 1.377 & 0.553 & 1.311 & 0.511 & 1.258 & 0.483 & 1.220 \\
IPCC & 0.635 & 1.387 & 0.584 & 1.305 & 0.507 & 1.252 & 0.454 & 1.203 \\
PMF & 0.568 & 1.537 & 0.487 & 1.321 & 0.451 & 1.221 & 0.433 & 1.171 \\
LACF & 0.682 & 1.500 & 0.650 & 1.468 & 0.610 & 1.416 & 0.582 & 1.381 \\
NCF & 0.440 & 1.325 & 0.385 & 1.283 & 0.372 & 1.253 & 0.362 & 1.205 \\
D2E-LF & 0.473 & 1.305 & 0.406 & 1.119 & 0.382 & 1.114 & 0.368 & 1.120 \\
LDCF & 0.413 & 1.310 & 0.385 & 1.255 & 0.362 & 1.204 & 0.346 & 1.185 \\
DFMI & 0.407 & 1.261 & 0.375 & 1.238 & 0.357 & 1.193 & 0.325 & 1.173 \\
NCRL & 0.358 & 1.262 & 0.338 & 1.222 & 0.321 & 1.217 & 0.307 & 1.130 \\
FRLN & 0.369 & 1.289 & 0.341 & 1.235 & 0.320 & 1.201 & 0.305 & 1.175 \\
\midrule
\textbf{PDS-Net} & \textbf{0.342} & \textbf{1.243} & \textbf{0.306} & \textbf{1.117} & \textbf{0.289} & \textbf{1.113} & \textbf{0.279} & \textbf{1.093} \\
\midrule
\textbf{Gains (\%)} & \textbf{7.32\%} & \textbf{3.57\%} & \textbf{10.26\%} & \textbf{9.55\%} & \textbf{9.69\%} & \textbf{7.33\%} & \textbf{8.52\%} & \textbf{6.98\%} \\
\bottomrule
\end{tabular}
\end{threeparttable}
\end{table*}

\begin{table*}[tp]
\centering
\caption{Performance Comparison with Different Training Ratios on Dataset D2 (Best Results in Bold Numbers)}
\label{tab:performance_comparison}
\begin{threeparttable}
\setlength{\tabcolsep}{3.5mm}
\begin{tabular}{c c c c c c c c c c c}
\toprule
\textbf{Method} & \multicolumn{2}{c}{\textbf{D2.1}} & \multicolumn{2}{c}{\textbf{D2.2}} & \multicolumn{2}{c}{\textbf{D2.3}} & \multicolumn{2}{c}{\textbf{D2.4}} & \multicolumn{2}{c}{\textbf{D2.5}} \\ 
\cmidrule(lr){2-3} \cmidrule(lr){4-5} \cmidrule(lr){6-7} \cmidrule(lr){8-9} \cmidrule(lr){10-11}
& MAE & RMSE & MAE & RMSE & MAE & RMSE & MAE & RMSE & MAE & RMSE \\ 
\midrule

UPCC & 0.555 & 1.317 & 0.542 & 1.022 & 0.466 & 0.820 & 0.428 & 0.787 & 0.389 & 0.754 \\
IPCC & 0.596 & 1.342 & 0.520 & 1.121 & 0.473 & 0.900 & 0.437 & 0.838 & 0.427 & 0.829 \\
UIPCC & 0.584 & 1.329 & 0.506 & 1.075 & 0.462 & 0.878 & 0.427 & 0.820 & 0.415 & 0.808 \\
HMF & 0.349 & 0.674 & 0.277 & 0.579 & 0.261 & 0.551 & 0.256 & 0.532 & 0.256 & 0.526 \\
D2E-LF & 0.653 & 1.638 & 0.633 & 1.577 & 0.607 & 1.564 & 0.600 & 1.563 & 0.590 & 1.556 \\
NCRL & 0.312 & 0.933 & 0.252 & 0.772 & 0.221 & 0.712 & 0.201 & 0.642 & 0.182 & 0.650 \\
CMF & 0.327 & 0.657 & 0.294 & 0.605 & 0.250 & 0.536 & 0.231 & 0.496 & 0.205 & 0.461 \\
LDCF & 0.234 & 0.585 & 0.199 & 0.548 & 0.183 & 0.509 & 0.172 & 0.479 & 0.161 & 0.472 \\
DFMI & 0.195 & 0.645 & 0.165 & 0.572 & 0.147 & 0.512 & 0.135 & 0.496 & 0.130 & 0.455 \\
FRLN & 0.185 & 0.635 & 0.162 & 0.562 & 0.136 & 0.522 & 0.131 & 0.443 & 0.131 & 0.458 \\
\midrule
\textbf{PDS-Net} & \textbf{0.170} & \textbf{0.518} & \textbf{0.144} & \textbf{0.464} & \textbf{0.129} & \textbf{0.431} & \textbf{0.121} & \textbf{0.396} & \textbf{0.119} & \textbf{0.395} \\
\midrule
\textbf{Gains (\%)} & \textbf{8.11\%} & \textbf{18.42\%} & \textbf{11.11\%} & \textbf{17.44\%} & \textbf{5.15\%} & \textbf{17.45\%} & \textbf{7.63\%} & \textbf{10.60\%} & \textbf{9.16\%} & \textbf{13.75\%} \\
\bottomrule
\end{tabular}
\end{threeparttable}
\vspace{-1em}
\end{table*}

\subsection{Evaluation Metrics}
In the field of QoS prediction, the primary criterion for evaluating the effectiveness of a model is prediction accuracy. The accuracy of a model is often measured using two metrics. The first metric is the mean absolute error function (MAE)\cite{chowdhury2020cahphf}, which is defined as:
\begin{equation}
\centering
MAE=\frac{\left ( \sum_{i,j}^{} \left |y_{i,j}-\hat{y}_{i,j} \right |_{abs}\right )}{ N }
\end{equation}
where $y_{i,j}$ is the real value and $\hat{y}_{i,j}$ is the predicted value of QoS property, and N is the number of QoS records. The second metric is the root mean squared error (RMSE)\cite{chowdhury2020cahphf}:
\begin{equation}
\centering
RMSE=\sqrt{\tfrac{\left ( \sum_{i,j}^{} (y_{i,j}-\hat{y}_{i,j} )^{2}\right ) }{ N }}
\end{equation}
In the field of QoS prediction, prediction accuracy is a crucial criterion for evaluation. Typically, two metrics are employed to measure accuracy. The first metric is the mean absolute error (MAE) function, which calculates the absolute difference between the labels and predicted values. The second metric is the root mean square error (RMSE), which is sensitive to larger or smaller values and assigns a relatively high weight to outliers. Lower values for these two metrics indicate more accurate predictions.

\subsection{Baseline Methods}

For ease of presentation, we compare PDS-Net with the following methods:


\begin{itemize}[leftmargin=*, itemsep=3pt,topsep = 3pt,partopsep=3pt]
\item \textbf{IPCC} \cite{carlkadie1998empirical}: The IPCC method computes the similarity between two services by borrowing the thinking from the service-based collaborative filtering algorithm.
\item \textbf{UPCC} \cite{13}: The UPCC method computes the similarity between two users by borrowing the thinking from the user-based collaborative filtering algorithm.
\item \textbf{LACF} \cite{18}: This method uses location information of users and services for service recommendation. LACF proposes a location-aware collaborative filtering method by incorporating the locations of both users and services, focusing on users physically near the target user.
\item \textbf{LFM} \cite{26}: This method learns the latent features of users and services for QoS prediction using the QoS matrix. Matrix factorization models allow for the incorporation of additional information such as temporal effects, implicit feedback, and confidence levels.
\item \textbf{PMF} \cite{41}: This method uses probabilistic factors in matrix factorization for service recommendation. PMF models both the generative process for the data and the missing data mechanism and improves performance by jointly learning these two models to predict ratings and model the data observation process.
\item \textbf{JCM} \cite{40}: This method uses a CNN to learn the features of neighbors and forms a feature matrix for QoS prediction. JCM is capable of inferring the user features or service features by using the learned deep latent features of neighbors and contains a novel similarity computation method.
\item \textbf{NCF} \cite{20}: This is a deep learning approach that combines MLP and MF for QoS prediction. NCF can express and generalize matrix factorization under its framework by leveraging a multi-layer perceptron to learn the user-item interaction function.
\item \textbf{CMF} \cite{77}: This is an outlier-resilient QoS prediction method. CMF removes outliers from QoS data using the isolated forest algorithm and utilizes Cauchy loss to measure the discrepancy between the observed QoS values and the predicted ones.

\item \textbf{LDCF} \cite{21}: This is an advanced deep learning approach with location-awareness for QoS prediction. LDCF proposes a new deep CF model with a similarity adaptive corrector (AC) in the output layer to correct the predictive quality of service.
\item \textbf{D2E-LF} \cite{wu2022double}: D2E-LF is the latest QoS prediction method based on machine learning. It provides a fast CF-based QoS prediction method.
\item \textbf{NCRL} \cite{zou2022ncrl}: NCRL is the QoS prediction method based on deep learning, which performs high-precision QoS prediction by fine-grained learning of user and service characteristics respectively.
\item \textbf{DFMI} \cite{zhang2023deep}: The method designs a feature mapping and inference network to obtain a high-dimensional feature matrix, and then, introduces a feature compensation block to compensate for potential feature information loss in the feature mapping and inference.
\item \textbf{FRLN} \cite{zou2024frln}:  FRLN is a state-of-the-art QoS prediction framework leveraging federated learning and a Residual Ladder Network to ensure data privacy while effectively capturing complex user-service relationships for precise QoS predictions.

\end{itemize}

In accordance with standard evaluation practices, outliers were removed when computing the MAE and RMSE metrics for all methods applied to dataset D2. 
However, it should be noted that some methods were specifically designed and optimized for certain datasets, such as CMF, which is tailored to be resilient to outliers. 
As such, we only present the experimental results of CMF on dataset D2, as it cannot be reasonably applied to dataset D1. 
Furthermore, to ensure a fair comparison, each method was executed 10 times and the average results were reported.
\section{EXPERIMENTS}\label{sec:EXPERIMENTS}

\subsection{Performance Comparison}
\noindent
\textbf{Approach:} In order to evaluate the performance of our proposed approach, we compared it with several baselines, including traditional algorithms and state-of-the-art methods. We selected LDCF as one of the baselines, which is the latest QoS prediction method based on collaborative filtering of location features. In this experiment, we used the default configuration provided in its original paper. Another baseline is CMF, which is the latest outlier-based QoS prediction method and provides this experimental dataset. We used the default parameters provided in the original paper for CMF. We also included FRLN as a baseline, which is the latest DL-based QoS prediction method. 

The experimental results in Tables 3 and 4 underscore the superior performance of PDS-Net over traditional and state-of-the-art QoS prediction methods, including the latest FRLN approach, across various datasets. In Dataset D1, PDS-Net achieves an average improvement of 8.2\% over all baselines, demonstrating its ability to effectively capture complex user-service relationships. Specifically, compared to FRLN, PDS-Net achieves notable gains in MAE of 7.32\%, 10.26\%, 9.69\%, and 8.52\% across different scenarios, along with improvements in RMSE of 3.57\%, 9.55\%, 7.33\%, and 6.98\%. These results highlight the efficacy of PDS-Net’s feature extraction mechanisms, which are capable of learning both low-dimensional and high-dimensional latent features, making it especially effective in diverse QoS prediction environments.

In Dataset D2, which includes the RT dataset with outliers, PDS-Net achieves substantial improvements over FRLN and other methods, with gains of 8.11\%, 11.11\%, 5.15\%, 7.63\%, and 9.16\% in MAE, and 18.42\%, 17.44\%, 17.45\%, 10.60\%, and 13.75\% in RMSE. These results underscore PDS-Net's robustness in handling data with outliers, while FRLN’s effectiveness is limited in this scenario due to sparse data density. On the QoS dataset without outliers, PDS-Net efficiently learns high-dimensional latent distributions, achieving prediction accuracy comparable to CMF but using only 10\% of the training data, demonstrating its superior data efficiency and prediction

\noindent
\textbf{Results:} The experimental results demonstrate that the proposed approach outperforms all baselines in terms of both MAE and RMSE. This suggests that the proposed approach is more accurate than the other methods tested. 

\subsection{The impact of probabilistic supervision}

\noindent
\textbf{Approach:} To evaluate the effectiveness of deep probabilistic supervision module, we conduct three comparative experiments as follows:\\
a) PDS: PDS-Net employs probabilistic supervision.\\
b) PDS w/o P: PDS-Net employs a supervision network without probabilistic spaces.\\
c) PDS w/o DS: PDS-Net employs a probabilistic network without supervision.\\
In experiment b), we use the standard deep supervised network structure. The depth features are passed through the decoder and then supervised with the real label. Experiment c) refers to the scenario where PDS-Net no longer utilizes the uncertain deep feature $Z_1$, which is obtained by sampling from $P_1$.

\begin{figure} 
\centering 

\subfigtopskip=2pt 
\subfigbottomskip=2pt 

\subfigure[$P_1$]{\includegraphics[width=0.49\linewidth]{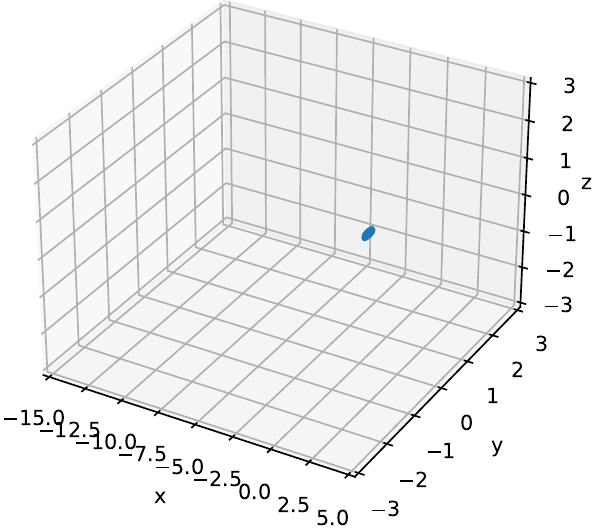}}
\subfigure[$P_r$]{\includegraphics[width=0.49\linewidth]{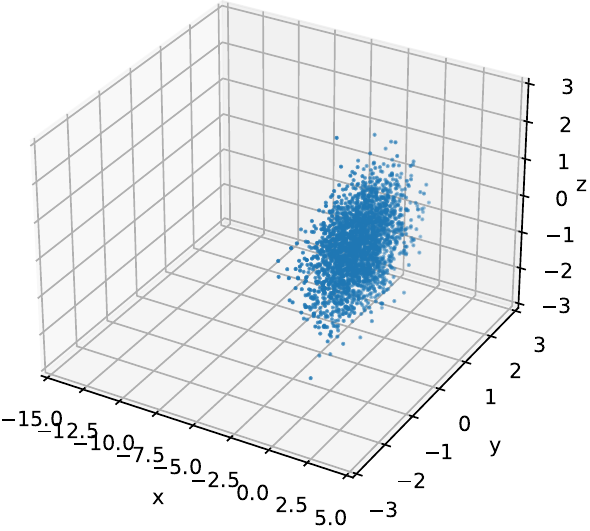}}
\subfigure[$P_1'$]{\includegraphics[width=0.49\linewidth]{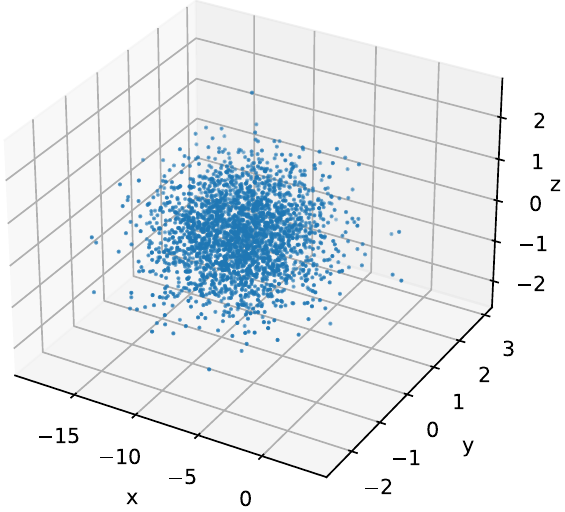}}
\subfigure[$P_2$]{\includegraphics[width=0.49\linewidth]{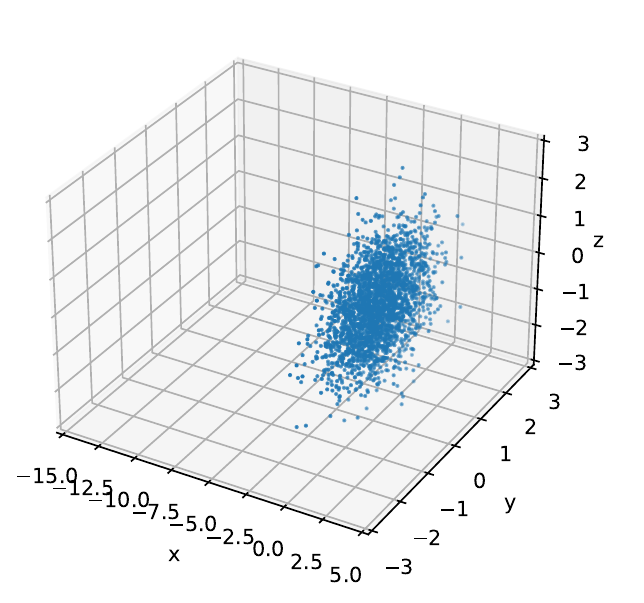}}
\caption{Probability distribution learning process.$P_1$ is the probability distribution learned from the missing features. $P_r$ is a probability distribution learned from ground truth labels. $P_1'$ is the distribution of p1 after training through $P_r$. $P_2$ is the probability distribution of deep features in the middle layer deep probabilistic supervision}
\label{fig:Gv}
\end{figure}

\noindent
\textbf{Visualization of the Probabilistic Deep Supervision Process:}
We present the visualization results of the training process using different distributions in the probabilistic supervised structure in \autoref{fig:Gv}. 
We selected D2.1 dataset and randomly sampled 2500 points from each distribution for visualization.  We picked a sample with missing features to visualize his training process.
\autoref{fig:Gv}(a) shows the visualization result of method c), where the network failed to effectively learn a distribution due to the lack of features, resulting in the convergence of features sampled from $P_1$ to a single point. 
It is observed that $P_1'$ has learned some distribution features from $P_r$ after PDS-network, resulting in dispersed sampling points instead of being concentrated in a single point. Finally, \autoref{fig:Gv}(d) presents the probability distribution of the middle layer, showing that $P_2$ becomes more similar to $P_r$ after being optimized by the middle layer. 
The MAE index of this method was 0.181. On the other hand, \autoref{fig:Gv}(b) displays the probability distribution learned by the supervised network from the true labels, while \autoref{fig:Gv}(c) shows the probability distribution $P_1'$ obtained after training with $P_r$. 

\begin{figure}
\centering 
\subfigtopskip=2pt 
\subfigbottomskip=2pt 
\subfigure[MAE]{\includegraphics[width=0.49\linewidth]{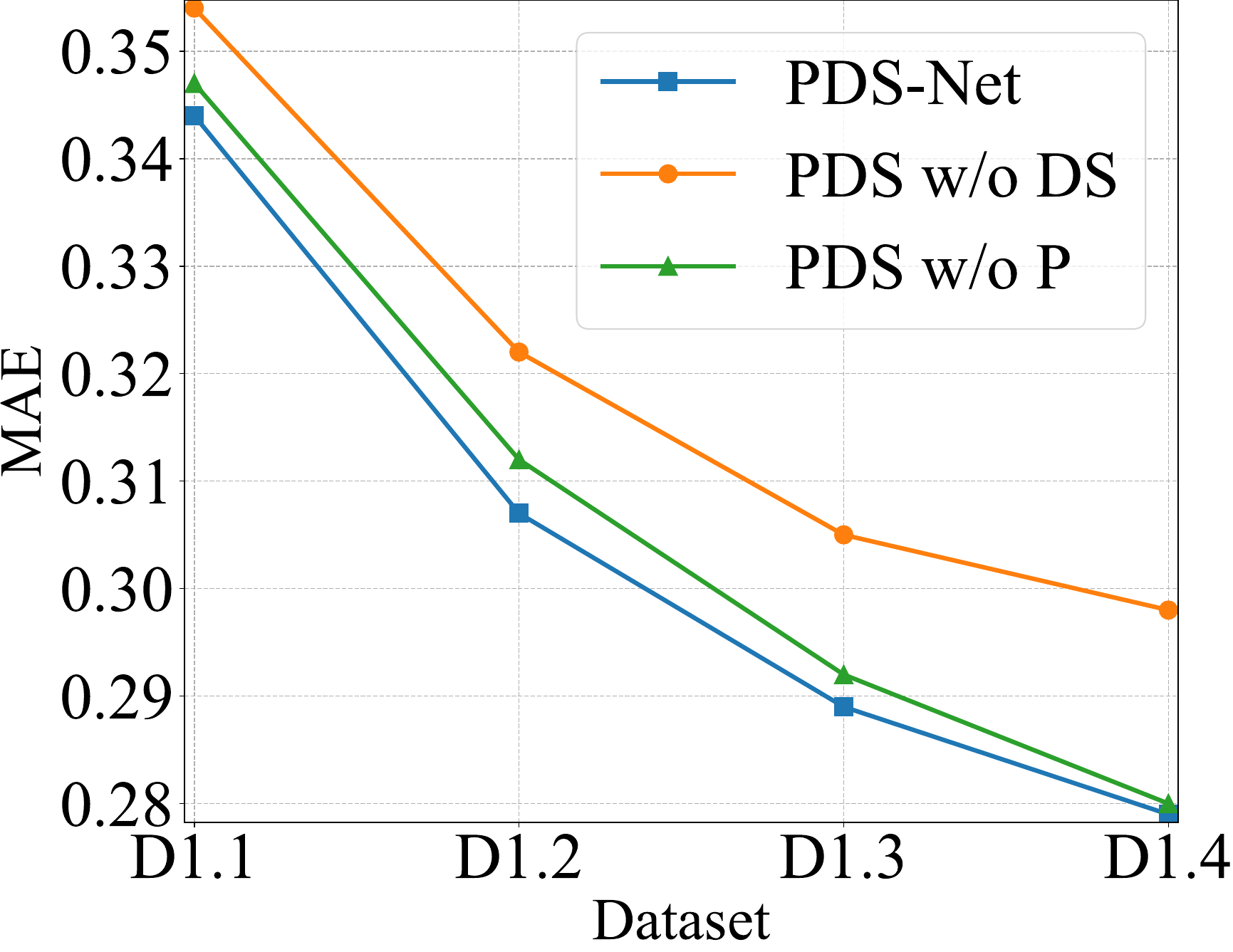}}
\subfigure[RMSE]{\includegraphics[width=0.49\linewidth]{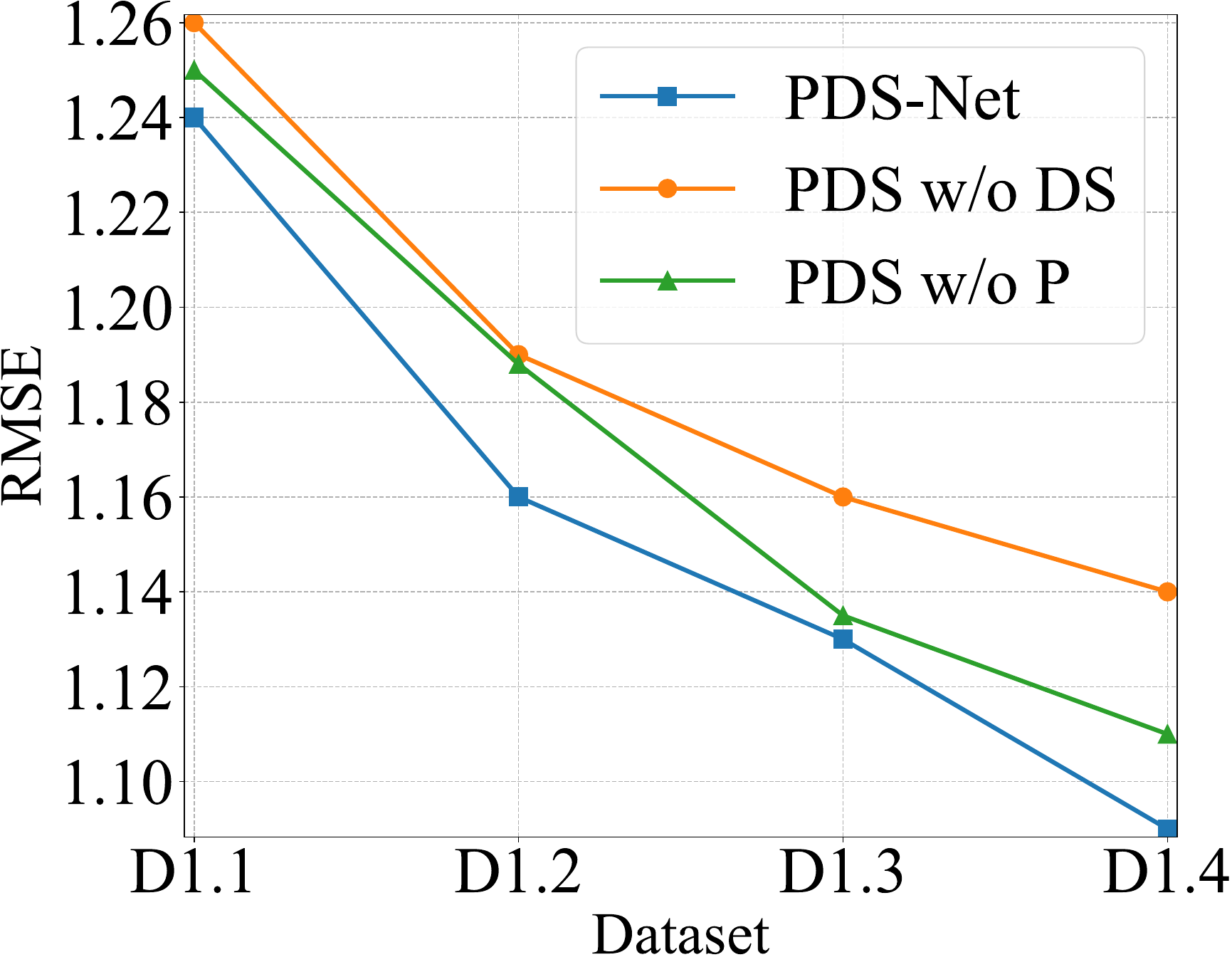}}
\subfigure[MAE]{\includegraphics[width=0.49\linewidth]{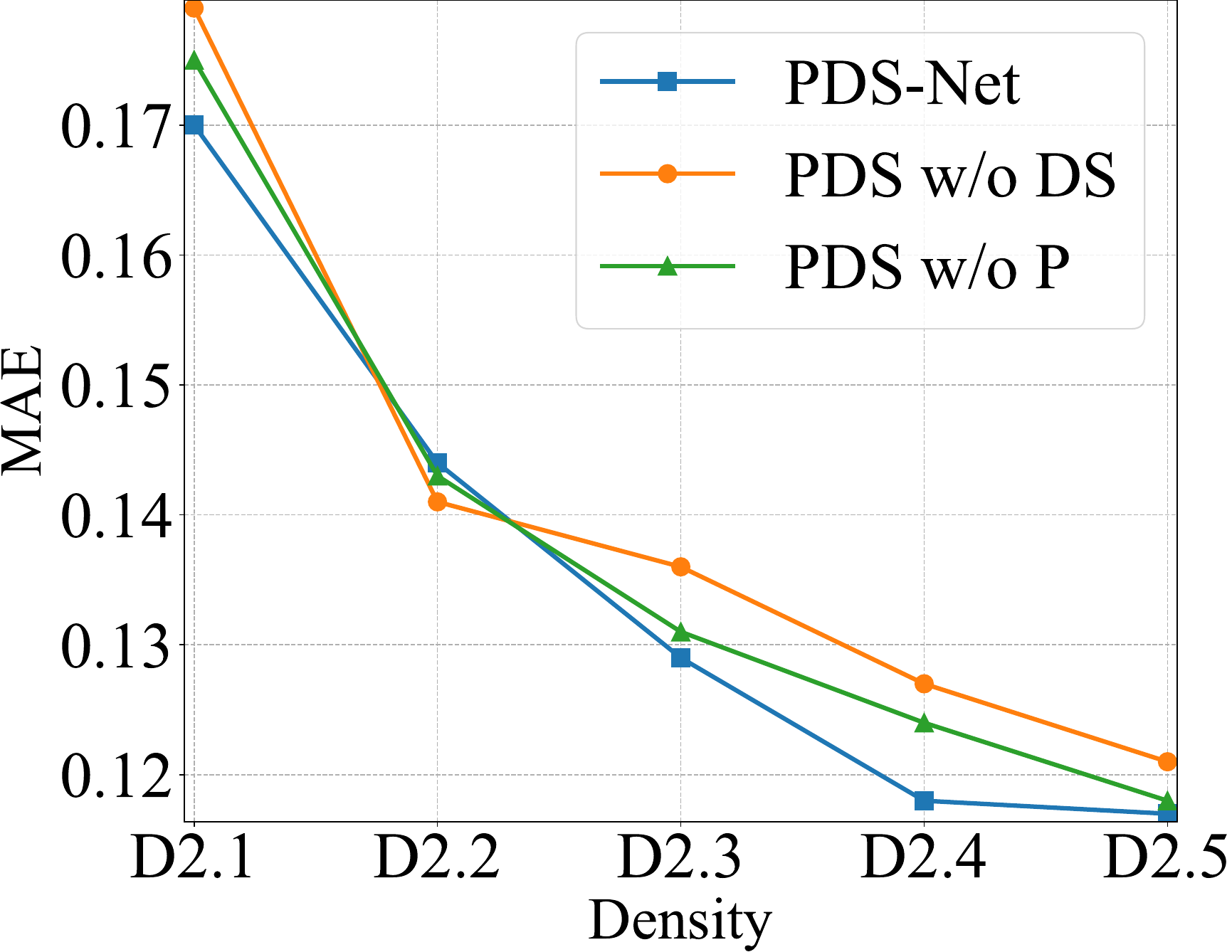}}
\subfigure[RMSE]{\includegraphics[width=0.49\linewidth]{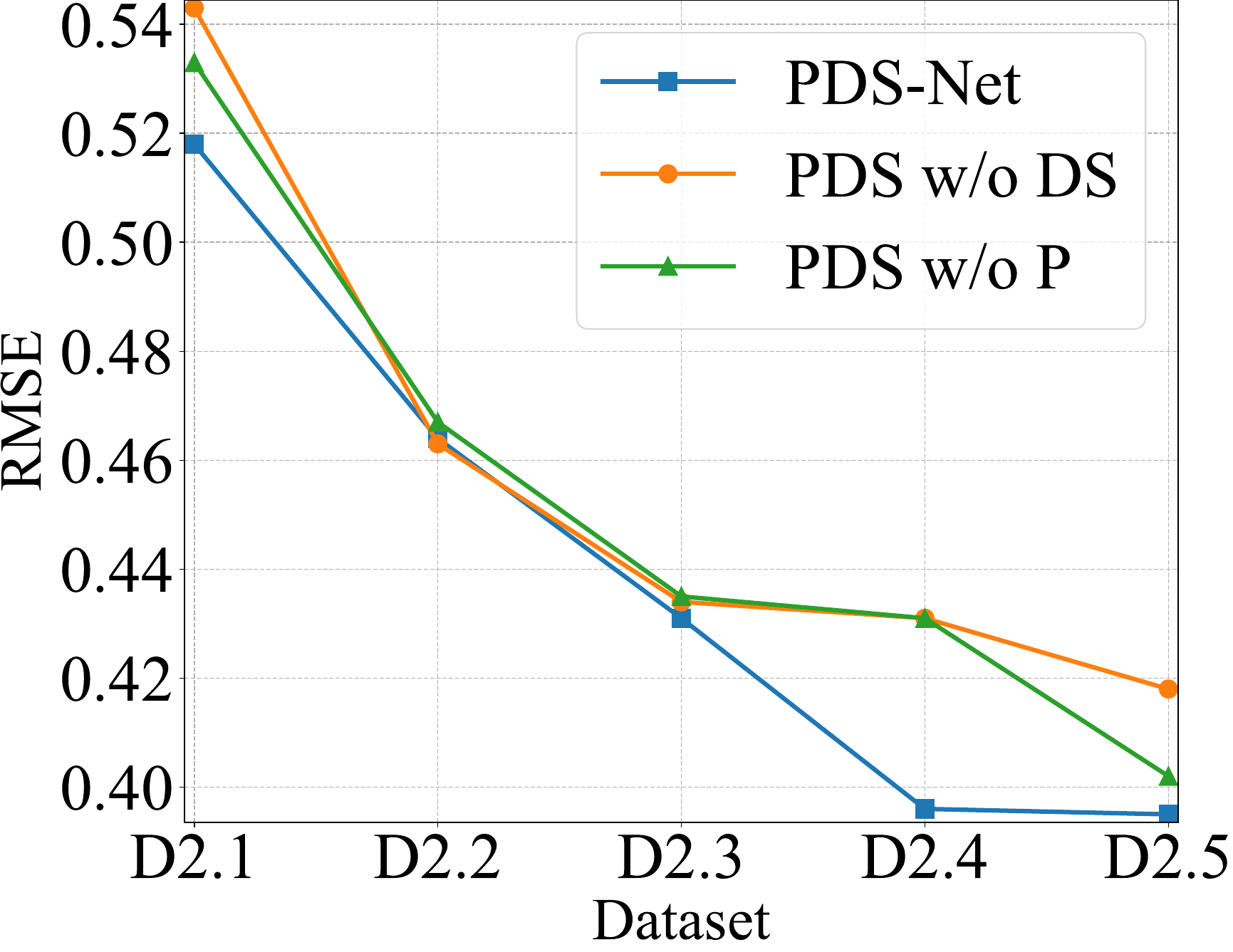}}
\caption{The impact of probabilistic supervision module}

\label{fig:prosup}
\vspace{-1em}
\end{figure}
The superiority of PDS-Net performance over other methods is demonstrated in \autoref{fig:prosup}. 
The results indicate that in datasets D1 and D2, the prediction performance of b) outperforms that of c), which suggests that the deeply supervised network performs better than probabilistic regression networks based on uncertain features. 
However, in datasets D2.1-D2.3, a), b), and c) exhibit similar performance, which may be due to the sparsity of the training data. Therefore, it can be concluded that c) is more advantageous than other methods.

\begin{figure} 
\centering 
\subfigtopskip=2pt 
\subfigbottomskip=2pt 
\subfigure[MAE]{\includegraphics[width=0.49\linewidth]{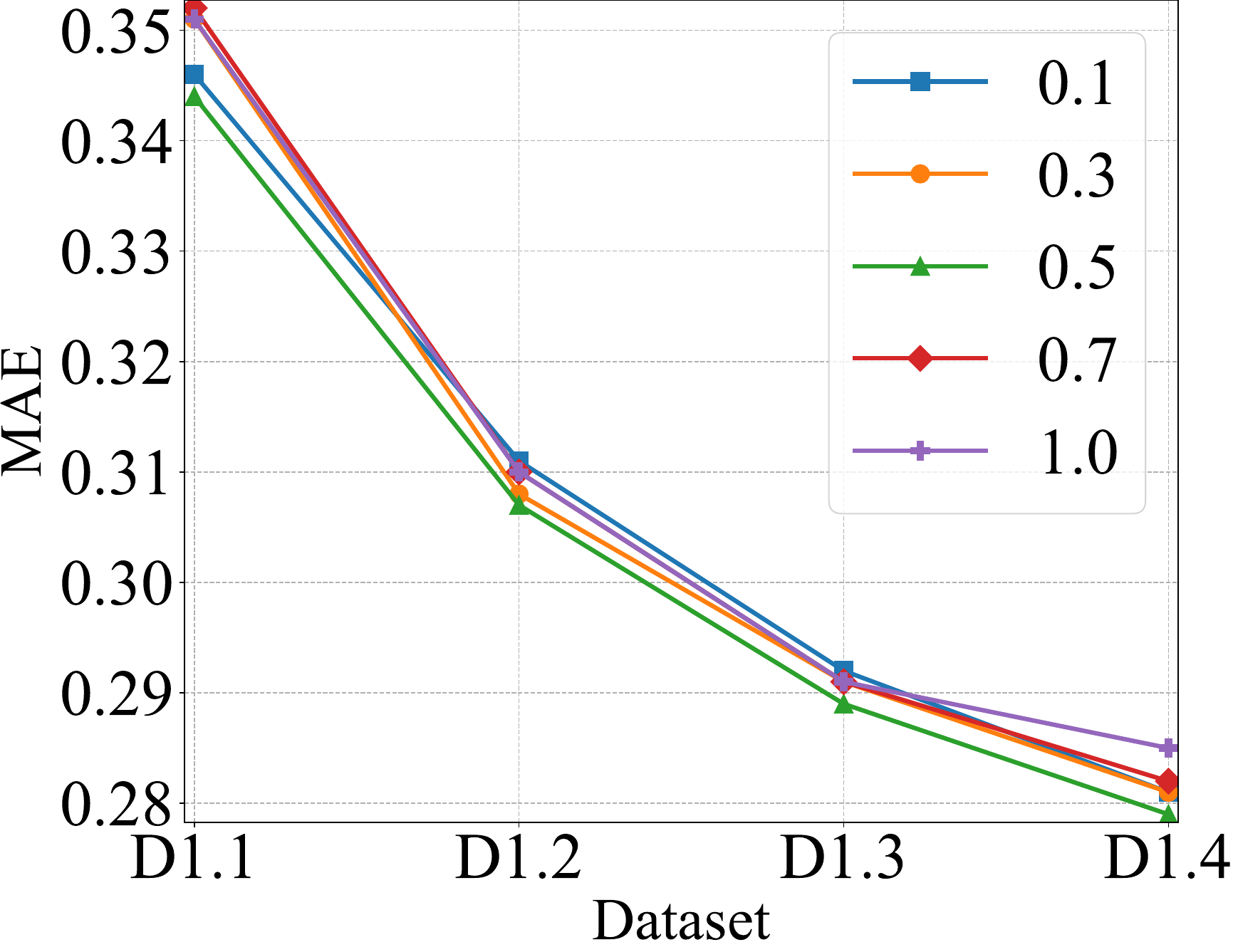}}
\subfigure[RMSE]{\includegraphics[width=0.49\linewidth]{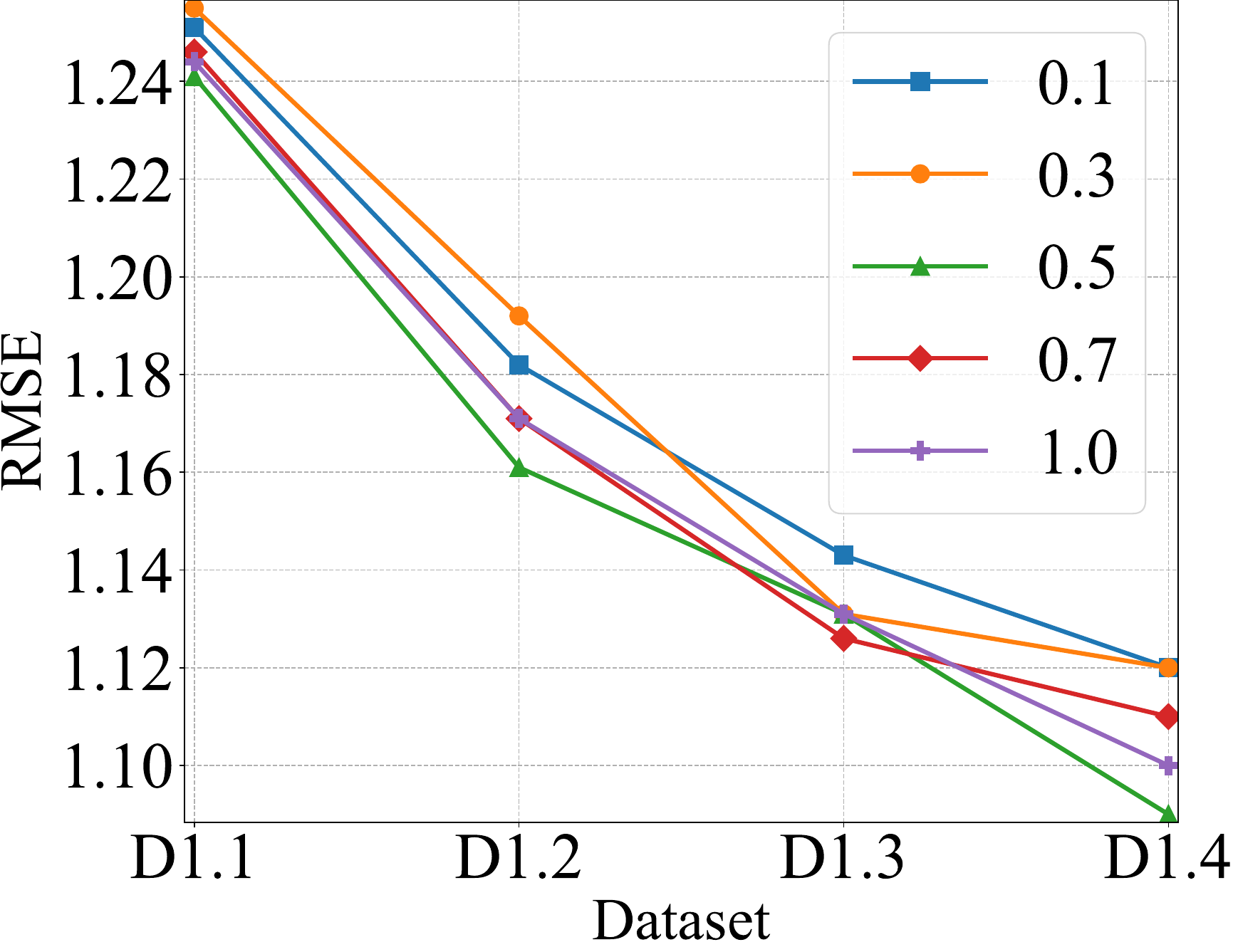}}
\subfigure[MAE]{\includegraphics[width=0.49\linewidth]{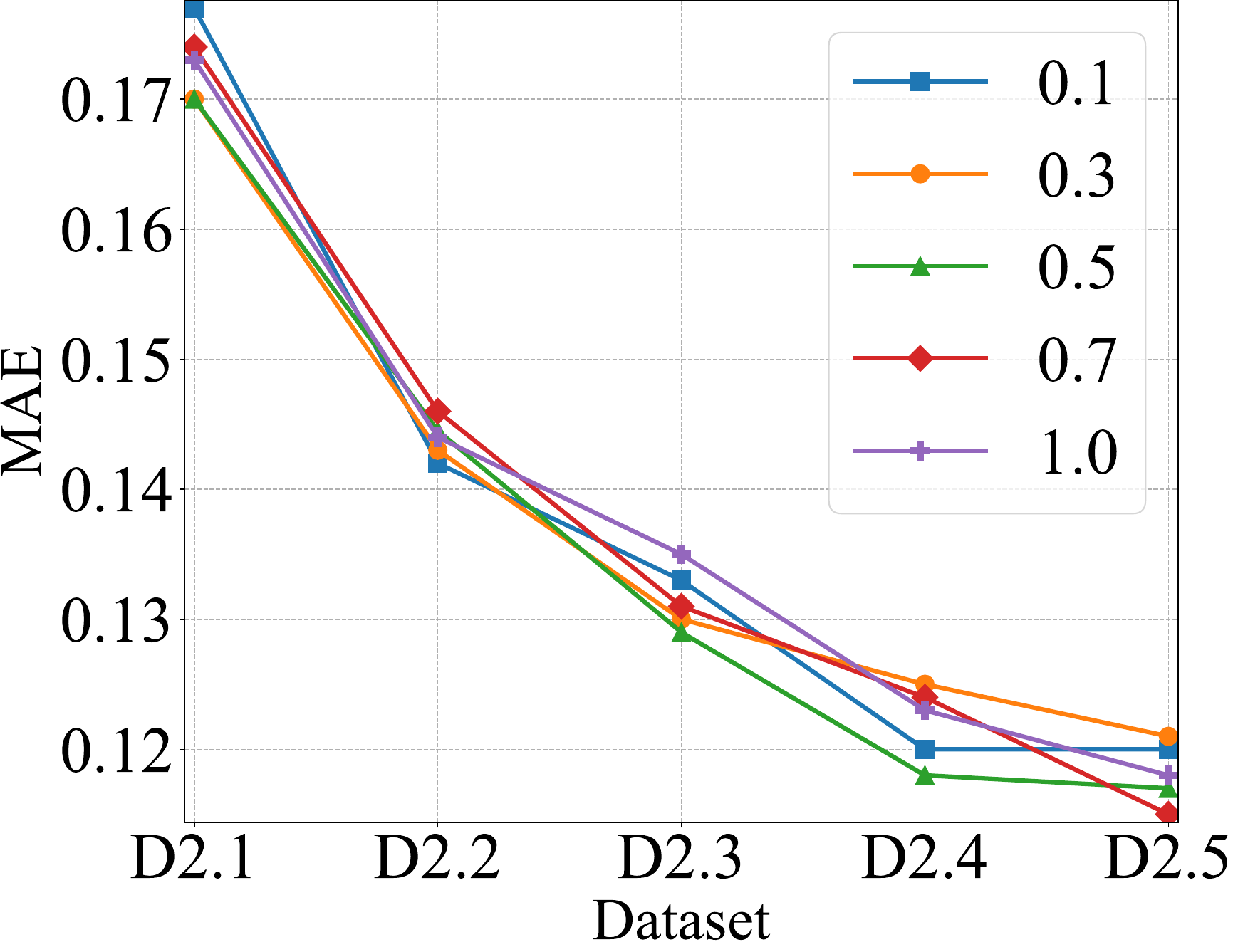}}
\subfigure[RMSE]{\includegraphics[width=0.49\linewidth]{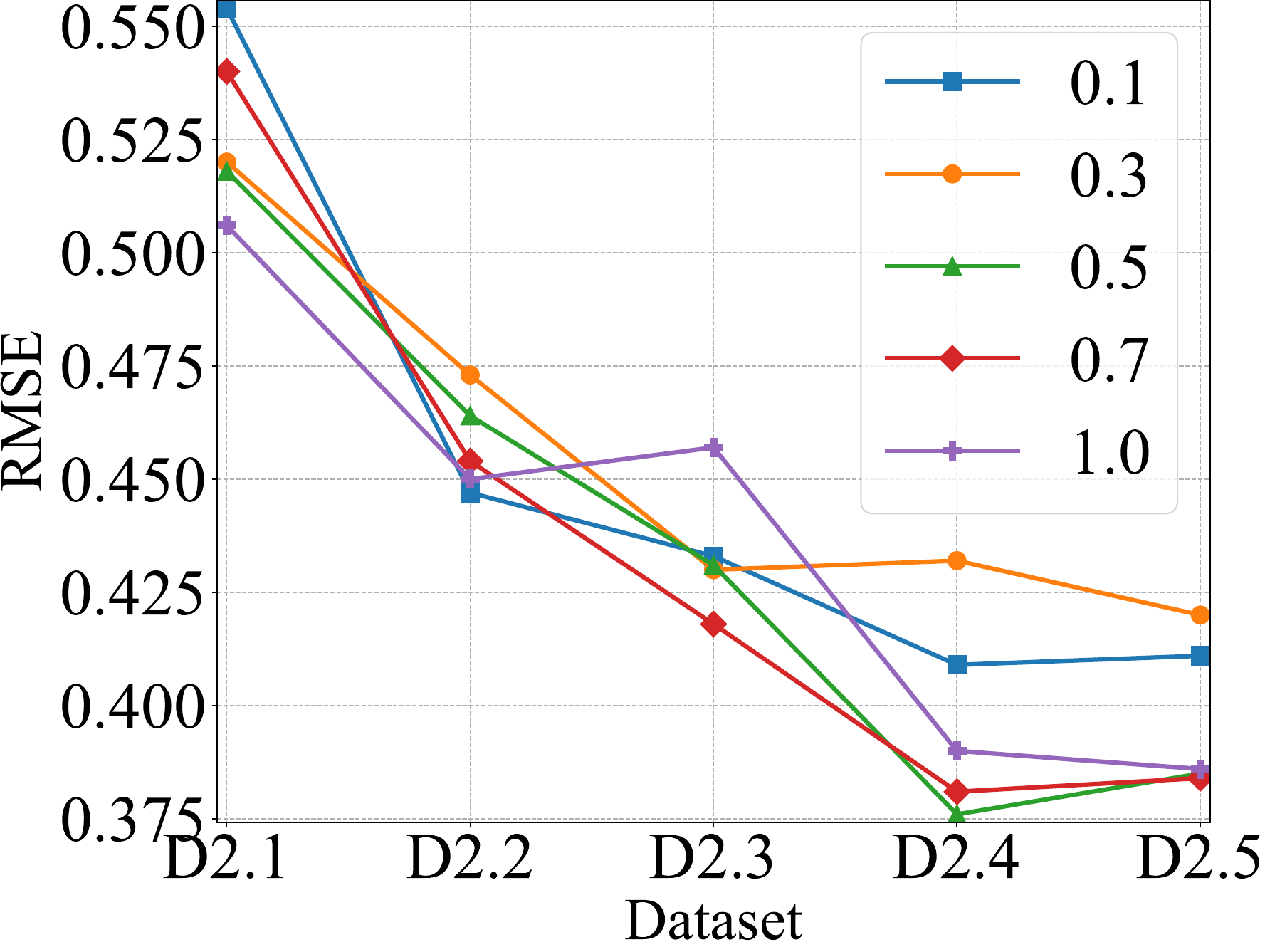}}
\caption{The results of PDS-Net with different cond in loss}
\vspace{-1em}
\label{fig:cond}
\vspace{-1em}
\end{figure}
\noindent
\textbf{Results:} Experiments show that the prior distribution is effectively changed by the posterior distribution during training. Moreover, the changed prior distribution can effectively improve the prediction accuracy.
\begin{figure} 
\centering 
\subfigtopskip=2pt 
\subfigbottomskip=2pt
\subfigure[MAE]{\includegraphics[width=0.49\linewidth]{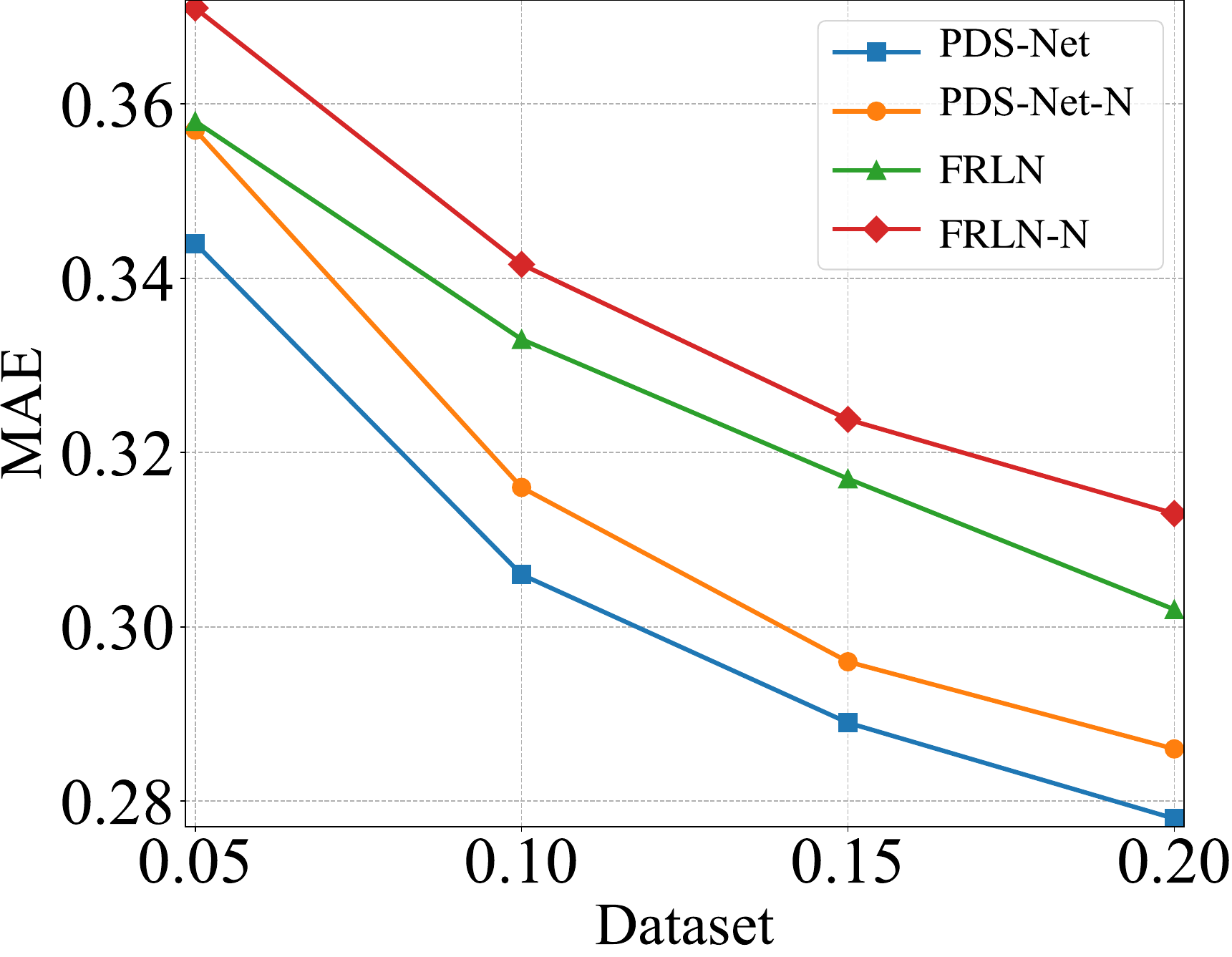}}
\subfigure[RMSE]{\includegraphics[width=0.49\linewidth]{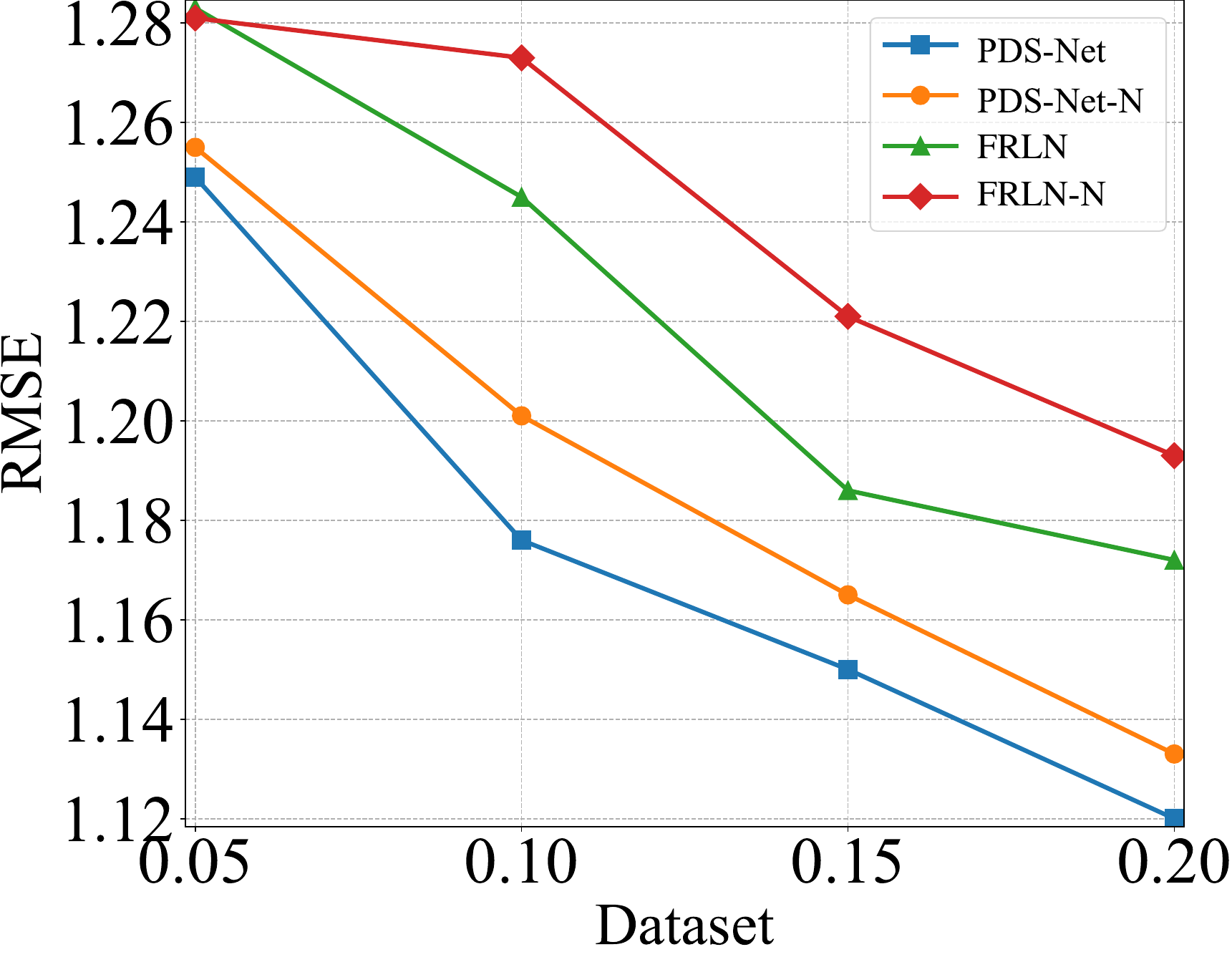}}
\caption{Prediction results  on incorrect features datasets}
\vspace{-1em}
\label{fig:noise}
\end{figure}
\vspace{-1em}
\subsection{The Impact of Incorrect Feature}

\noindent
\textbf{Approach:} Existing datasets lack labels to judge the wrong feature data.
To assess the impact of noise on model accuracy, we generated a noisy dataset based on D1. 
Specifically, false location information was provided for 10\% of 339 randomly selected users. 
The experimental setup was as follows:\\
a) PDS-Net: PDS-Net trained and tested on dataset D1.\\
b) PDS-Net-N: PDS-Net trained and tested on dataset D1 with noise.\\
c) FRLN: FRLN trained and tested on dataset D1. \\
d) FRLN-N: FRLN trained and tested on dataset D1 with noise.

As shown in \autoref{fig:noise}, PDS-Net outperformed the baseline on both the clean and noisy datasets.
However, the prediction performance of both PDS-Net and FRLN decreased in the presence of noise, indicating that noise can adversely affect prediction models. Notably, PDS-Net was less affected by noise than FRLN.

\noindent
\textbf{Results:} The experimental results demonstrate that PDS-Net is less affected by noise than the baseline method, and the presence of noise has a relatively minor impact on the prediction performance of PDS-Net.

\begin{figure} 
\centering 
\subfigtopskip=2pt 
\subfigbottomskip=2pt 
\subfigure[MAE]{\includegraphics[width=0.49\linewidth]{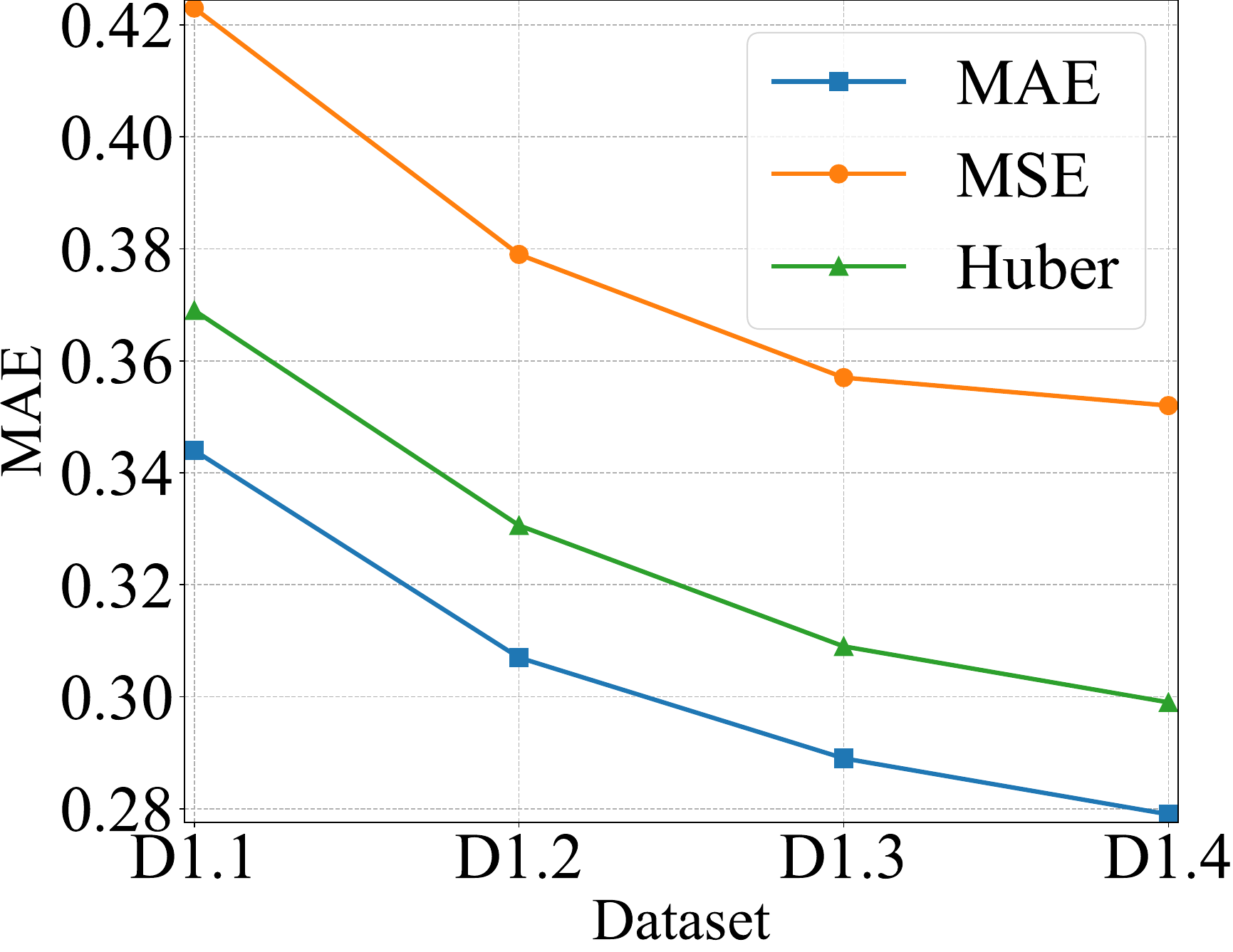}}
\subfigure[RMSE]{\includegraphics[width=0.49\linewidth]{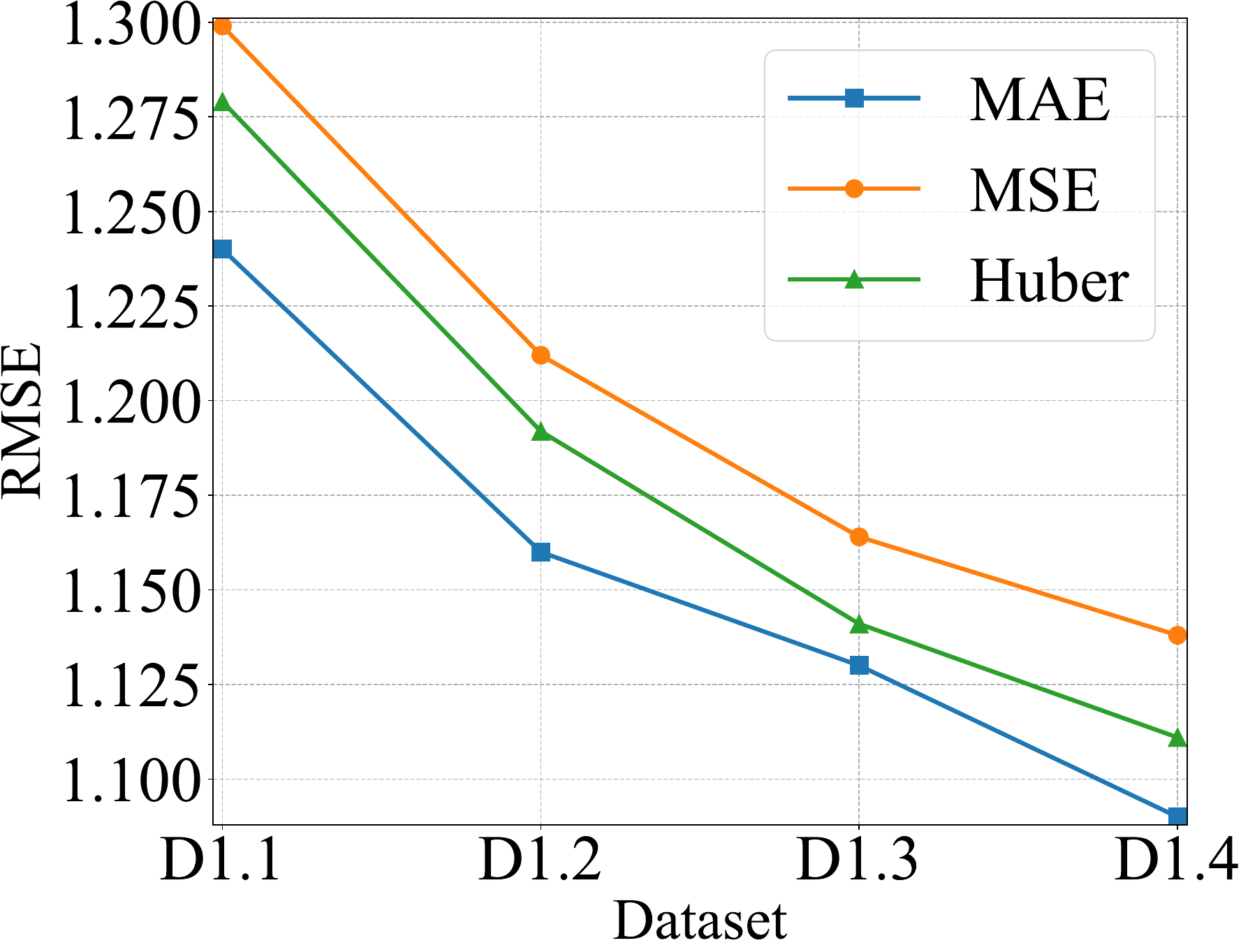}}
\subfigure[MAE]{\includegraphics[width=0.49\linewidth]{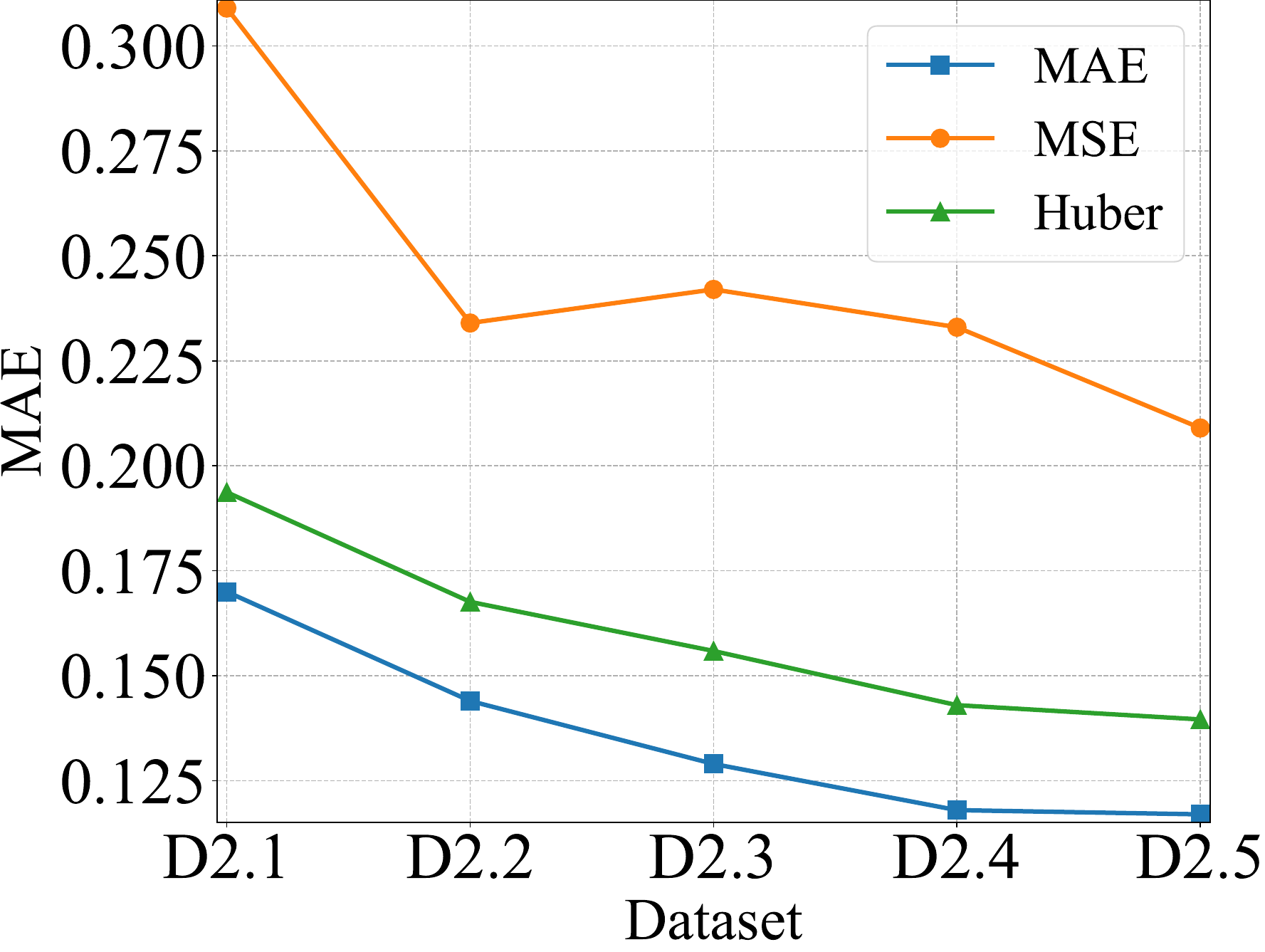}}
\subfigure[RMSE]{\includegraphics[width=0.49\linewidth]{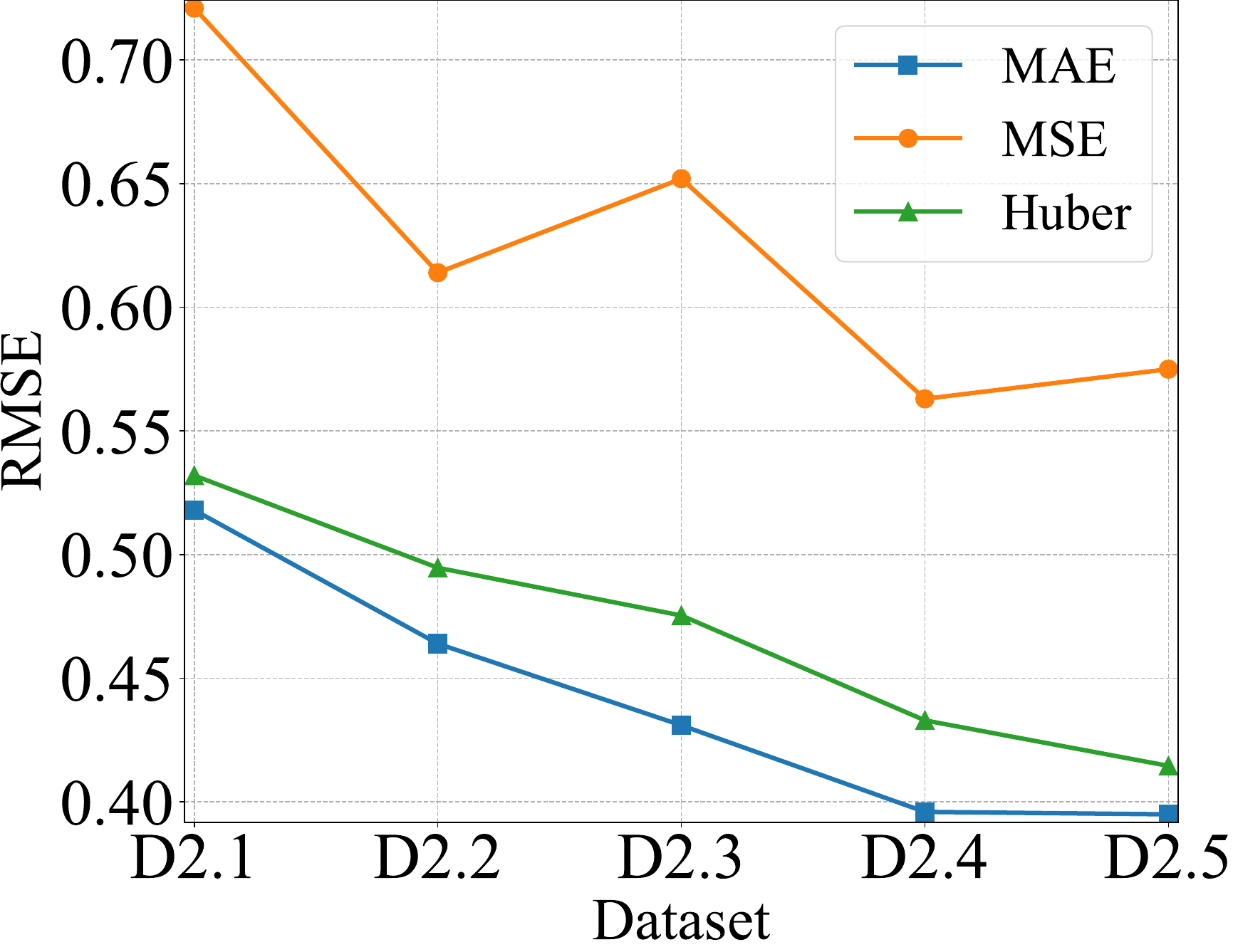}}
\caption{The results of PDS-Net with different loss function}
\vspace{-1em}
\label{fig:loss}
\vspace{-1em}
\end{figure}

\subsection{The Impact of Missing Feature}
\noindent
\textbf{Approach:} In order to better show the effectiveness of PDS-Net, we conduct experiments in the case of missing features.
In order to better show the robustness of the model in the case of missing features, we only keep the data items with missing feature data in the test set of dataset D1.
The experimental setup was as follows:\\
a) HSA-Net~\cite{86}: HSA-Net alleviates the feature missing problem by providing latent features through a probabilistic algorithm.\\
b) PDS-Net w/o PDS: Remove the probabilistic deep supervision process.\\
c) PDS-Net:Standard PDS-net process.\\
d) LDCF: LDCF is affected by location information, and the missing features may have a great impact on the method.\\
e) DFMI and FRLN: Recent deep learning methods are used for QoS prediction.

As shown in \autoref{tab:miss_comparison}, PDS-Net outperformed the baseline on both the datasets with miss features. However, the prediction performance of both PDS-Net and HSA-Net decreased in the presence of miss features, indicating that noise can adversely affect prediction models. 
This result is due to HSA-Net relying more on potential state features based on real labels, while LDCF relies more on known location information. This makes HSA-Net and PDS-Net equally more noise-resistant than LDCF.
Notably, PDS-Net was less affected by noise than HSA-Net. Specifically, the prediction performance of the backbone network of PDS-net is basically at the same level as the existing benchmarks. In data set D1, the prediction accuracy of missing features is much lower than the average level of the data set.


\begin{table}[ht]
    \centering
    \begin{threeparttable}
        \caption{Results of different approaches on dataset D1 with missing features}
        \label{tab:miss_comparison}
        \begin{tabular*}{\columnwidth}{@{\extracolsep{\fill}}lcccc@{}} 
            \toprule
            MAE & D1.1 & D1.2 & D1.3 & D1.4 \\
            \midrule
            LDCF & 0.467 & 0.414 & 0.401 & 0.389 \\
            DFMI & 0.486 & 0.433 & 0.401 & 0.383 \\
            FRLN & 0.474 & 0.435 & 0.398 & 0.382 \\
            HSA-Net & 0.419 & 0.393 & 0.378 & 0.367 \\
            PDS-Net w/o PDS & 0.417 & 0.396 & 0.374 & 0.363 \\
            PDS-Net & 0.407 & 0.382 & 0.364 & 0.355 \\
            \bottomrule
        \end{tabular*}
    \end{threeparttable}
    \vspace{-1em}
\end{table}
\subsection{The impact of different the task loss function}

\noindent
\textbf{Approach:} To evaluate the effectiveness of the feature distribution, we conducted four comparative experiments as follows:\\
a) MAE: Task loss function(Sec 3.5) is set to the mean absolute error (MAE).\\
b) MSE: Task loss function is set to the mean squared error (MSE).\\
c) Huber: Task loss function is set to the Huber loss function. 
The parameters of the Huber loss function are set according to the recommendation in \cite{21}.\\

The MAE and MSE loss functions are calculated using the standard TensorFlow calculations.


\autoref{fig:loss} shows that the performance of approach b) is significantly worse than that of a) and c). This can be attributed to the fact that most of the response times in the RT dataset are very small, while the response times for exceptions are very large. MSE loss is more sensitive to the exception response times, which affects the performance of b). On the other hand, MAE loss is less sensitive to anomalies than MSE loss and Huber loss.

\noindent
\textbf{Results:} Overall, The MAE loss performs better in various prediction scenarios, and therefore, we default to using it as the task loss function in this paper.

\begin{figure} 
\centering 
\subfigtopskip=2pt 
\subfigbottomskip=2pt 
\subfigure[MAE]{\includegraphics[width=0.494\linewidth]{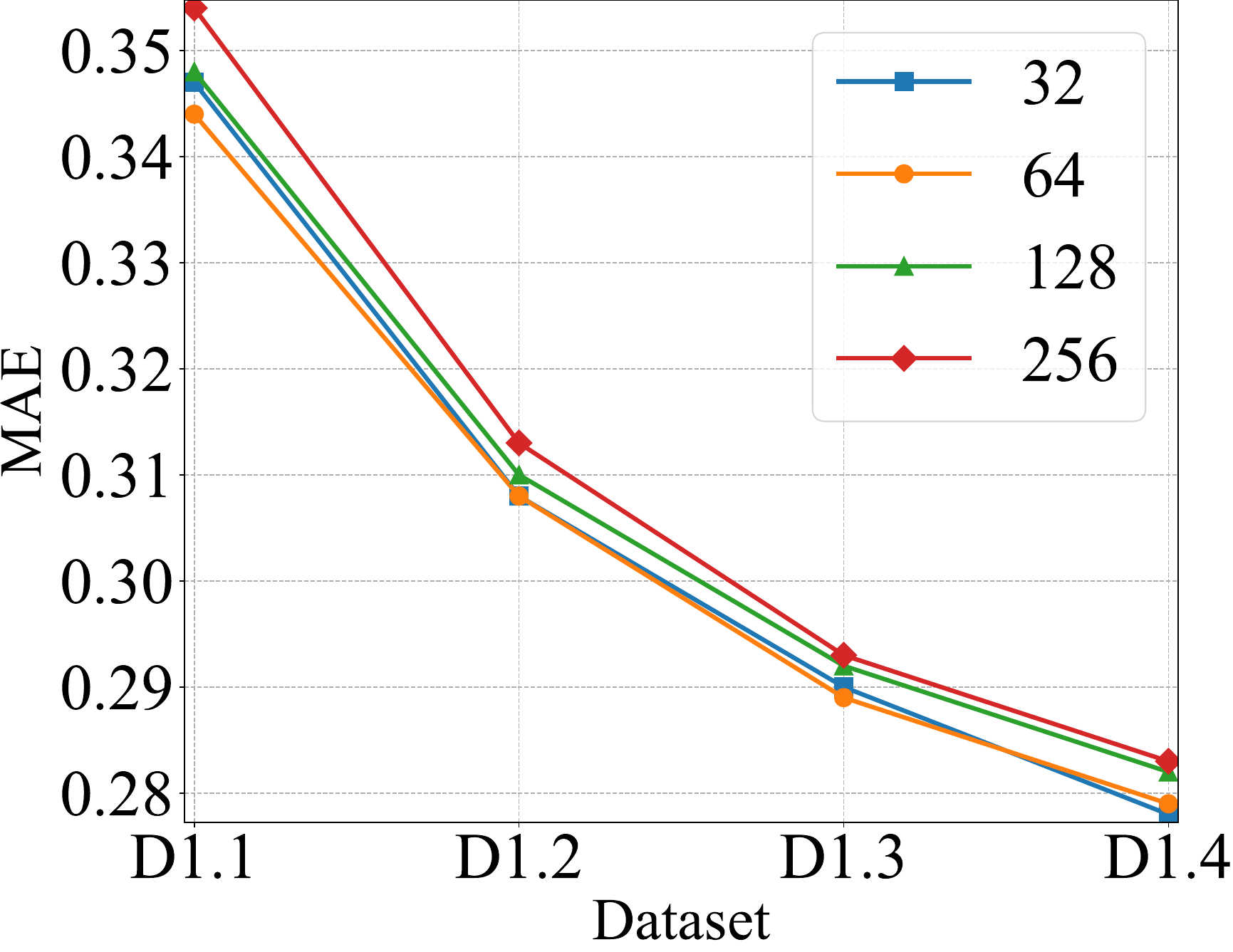}}
\subfigure[RMSE]{\includegraphics[width=0.494\linewidth]{pic/NMAE.pdf}}
\subfigure[MAE]{\includegraphics[width=0.494\linewidth]{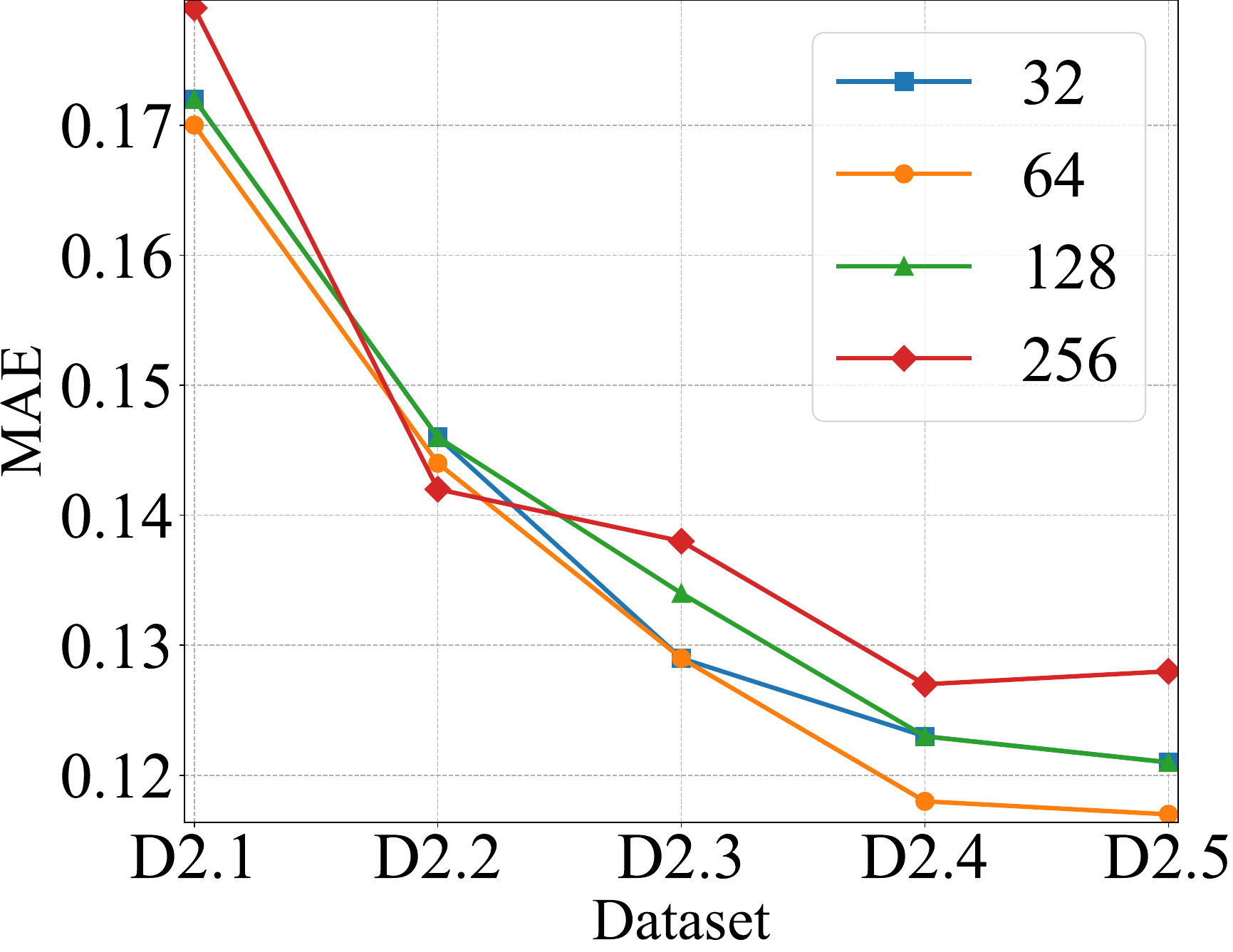}}
\subfigure[RMSE]{\includegraphics[width=0.494\linewidth]{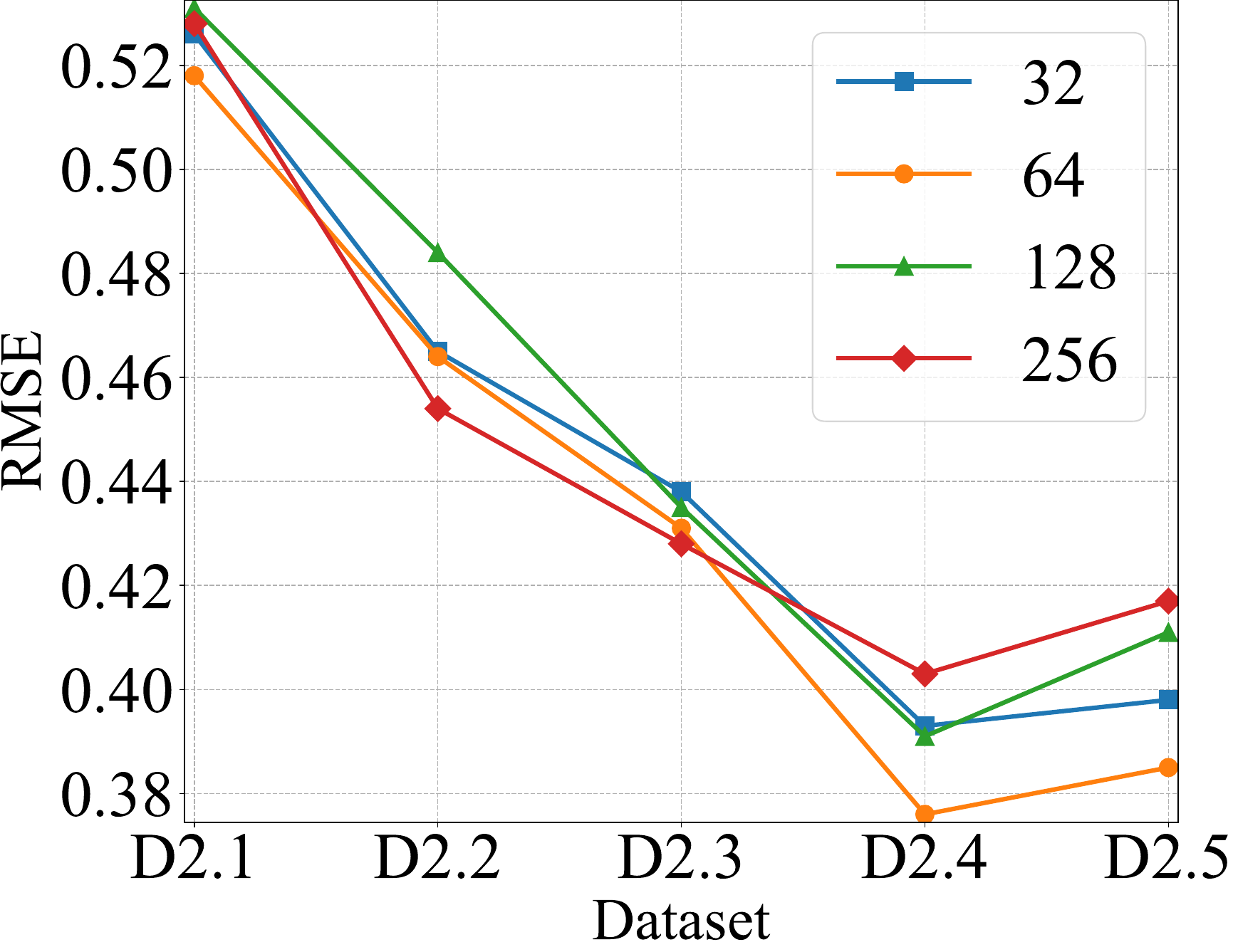}}
\caption{The results of PDS-Net with different N}
\vspace{-1em}
\label{fig:N}
\vspace{-1em}
\end{figure}

\subsection{Hyper-parameters Sensitivity Tests}

\emph{\textbf{1) The impact of different $\delta$}}

PDS-Net utilizes a task loss function with a hyperparameter ($\delta$) to optimize the main task or the probability space based on the absolute difference between the predicted value $\hat{y}$ and the true value $y$ of the backbone network.
Therefore, the value of ($\delta$ is crucial because it determines when our model considers the feature to be noisy.
The initial value of $\delta$ is initialized according to the value of MAE of previous methods.
According to LDCF, the MAE of QoS prediction method based on ResNet network is generally about 0.45. Therefore, the value of x is retrieved to both sides with an initial value of 0.5.
In particular, when the absolute difference is less than $\delta$, the model optimizes only the main task, and when the difference is too large, the model optimizes the probability space.

\textbf{Approach:} To evaluate the effectiveness of $\delta$, we conducted experiments with $\delta$ values of 0.1, 0.3, 0.5, 0.7, and 1.0.

As shown in \autoref{fig:cond}, the MAE and RMSE indicators perform best when $\delta$ is set to 0.5, with similar performance when $\delta$ is 0.7. The reason for this may be that when the value of $\delta$ is too small, the model focuses too much on optimizing the uncertain depth feature space, while when the value of $\delta$ is too large, the model's performance is similar to that of method b) in Section 5.2.

\textbf{Results:} In summary, the best prediction performance is achieved when $\delta=0.5$ for PDS-Net. Therefore, in this paper, we use $\delta=0.5$ as the default value.

\textbf{2) Impact of different parameter N}

As previously mentioned, PDS-Net establishes multiple Gaussian distributions with dimension N ($\mathbb{R}^{N}$), where N represents the complexity of the established Gaussian distribution and the dimensionality of the sampled features. By default, we sample deep features ($Z_1$ and $Z_2$) from Gaussian distribution ($P_1$ and $P_2$) using the default sampling method from Tensorflow\_probability.

\textbf{Approach:} To evaluate the effectiveness of N, we set the values of N to 32, 64, 128, and 256 for experiments. Specifically, when N=32, formulas 7-8 respectively obtain a vector of length 32 as the mean and standard deviation of the Gaussian distribution. At the same time, the sampling function defaults to sampling a vector of the same length from this distribution as the sampling feature.

\autoref{fig:N} shows that as the sample sizes increase, the RMSE and MAE increase when $N=64$. The MAE and RMSE indicators achieve the best performance when N is 64. The reason for this phenomenon may be that when N = 32, the low-dimensional deep features cannot represent the latent space well. When N is too large, PDS-Net cannot learn such a high-dimensional space based on the existing features and data.

\textbf{Results:}
When $N=64$, PDS-Net achieves good prediction performance in different situations. Therefore, the default value of $N$ in this paper is 64. This means that the shape of the variance and expectation of all Gaussian distributions in this paper is $\mathbb{R}^E$ and $\mathbb{R}^{E\times E}$, respectively.

\emph{\textbf{3) Impact of Different Embedded Dimensions (E)}}

\begin{figure} 
\centering 

\subfigtopskip=2pt 
\subfigbottomskip=2pt 

\subfigure[MAE]{\includegraphics[width=0.49\linewidth]{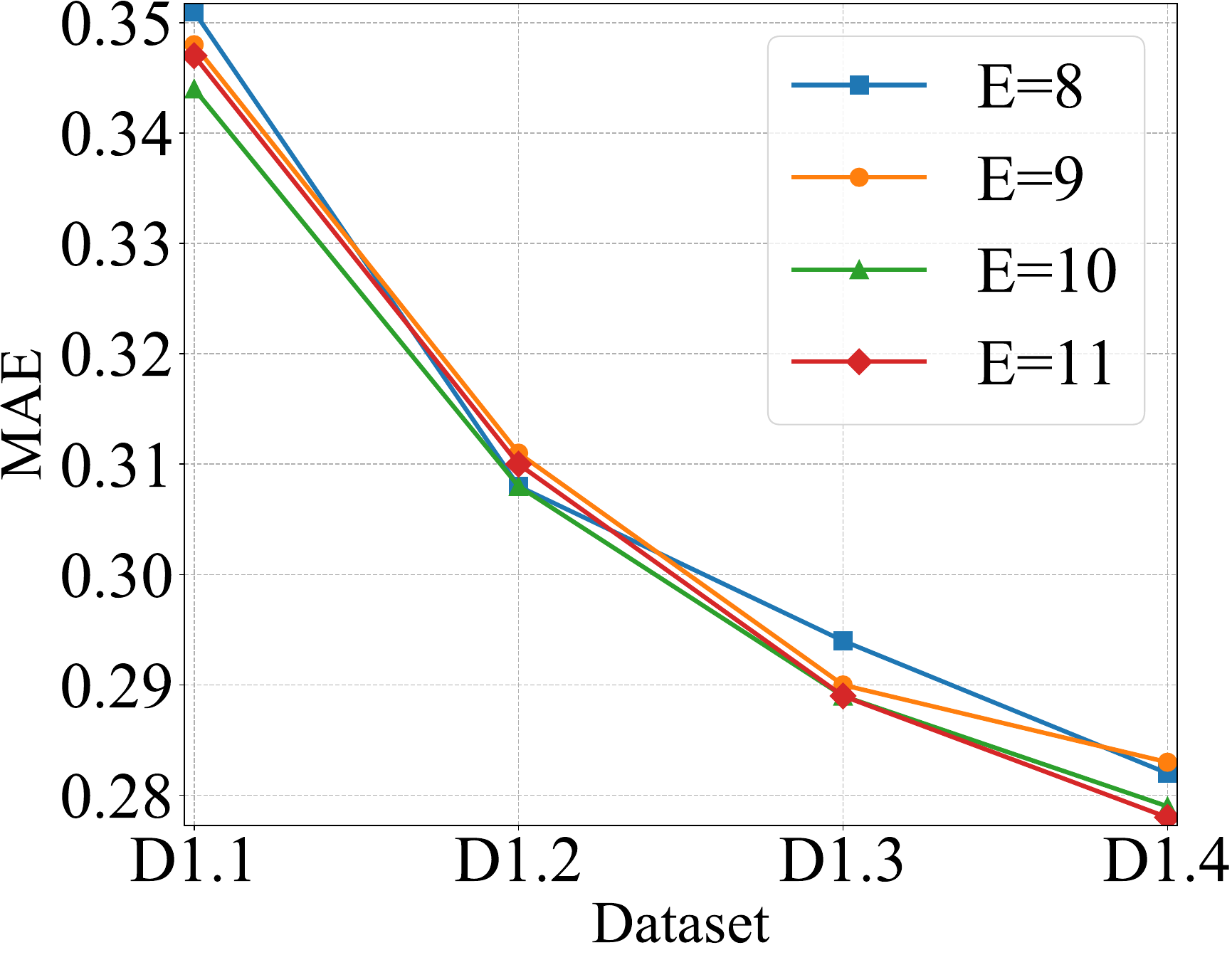}}
\subfigure[RMSE]{\includegraphics[width=0.49\linewidth]{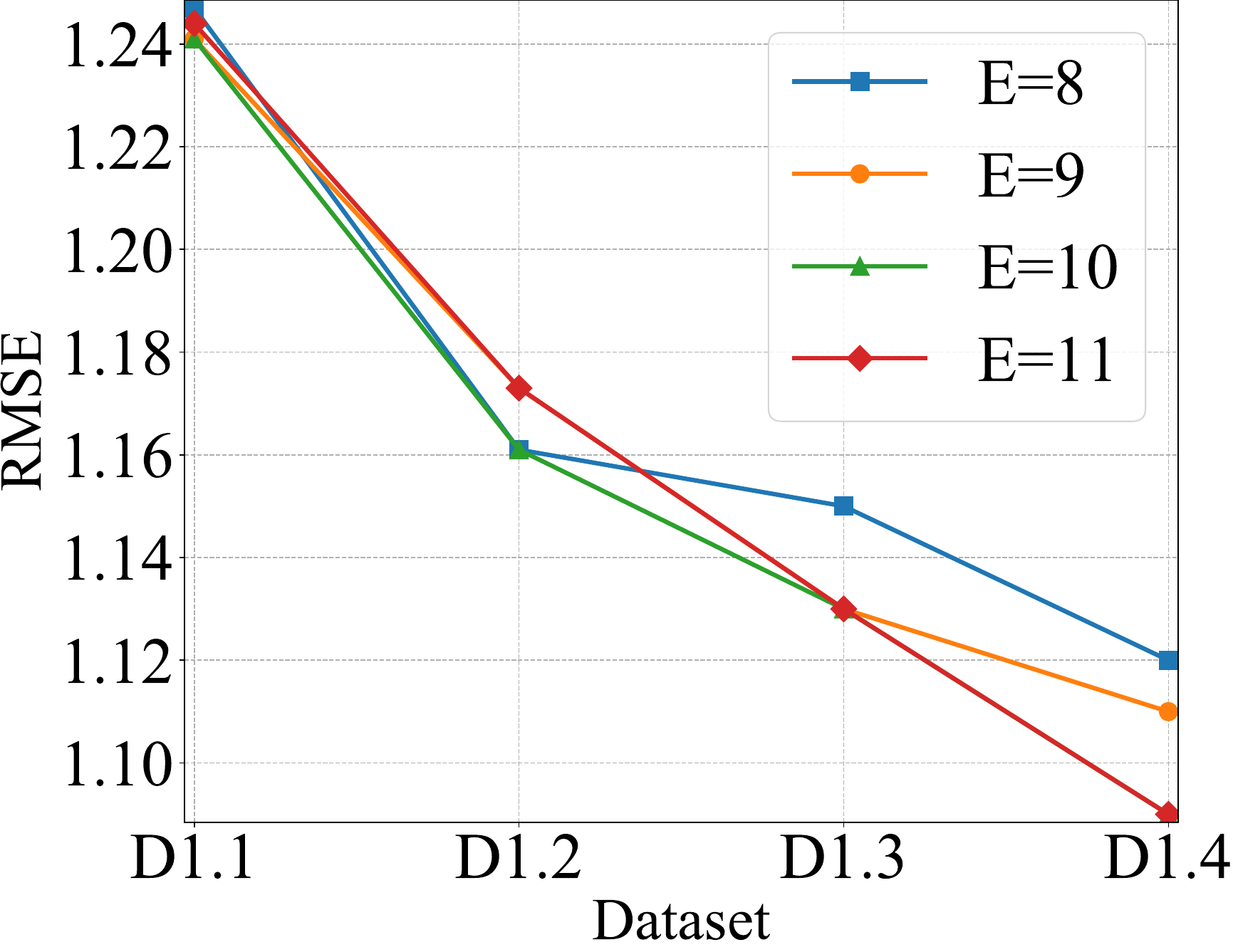}}
\subfigure[MAE]{\includegraphics[width=0.49\linewidth]{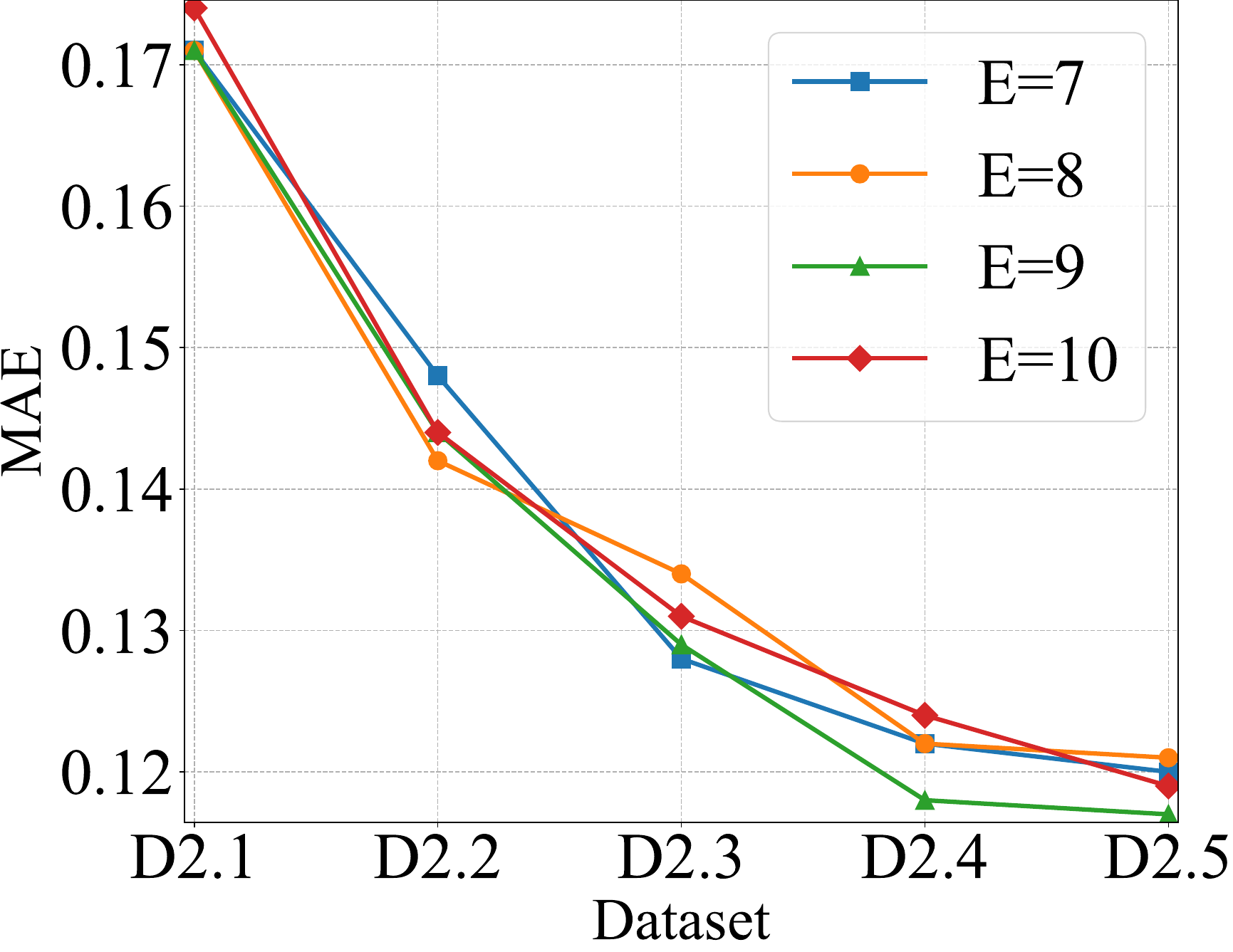}}
\subfigure[RMSE]{\includegraphics[width=0.49\linewidth]{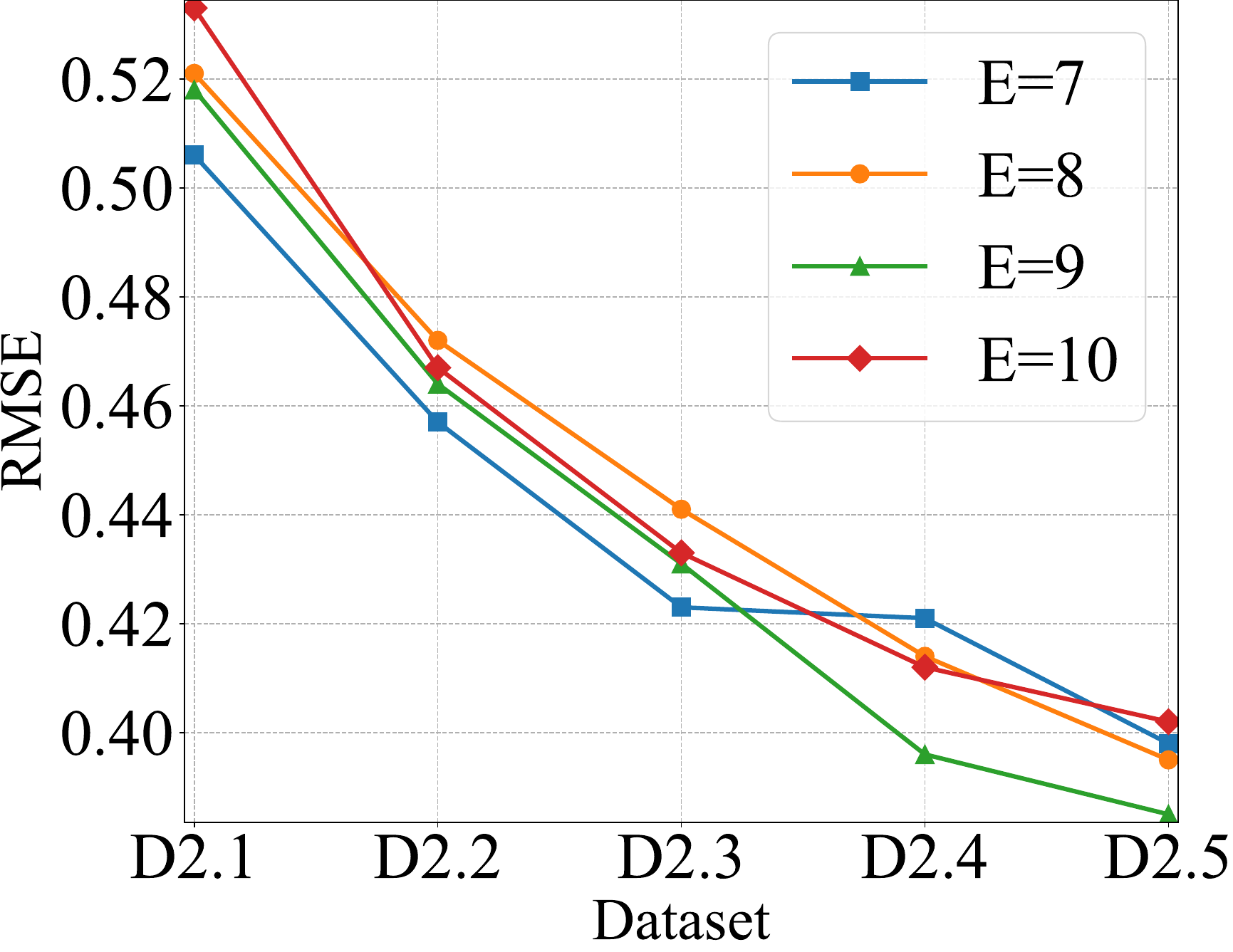}}
\caption{The results of PDS-Net with different E}
\vspace{-1em}
\label{fig:E}
\vspace{-1em}
\end{figure}
 In PDS-Net, known features are projected into a K-dimensional vector through the Keras embedding layer.

\noindent
\textbf{Approach:} To evaluate the effectiveness of E, we conducted experiments with different values of E, specifically 8, 9, 10, and 11. For the category of service ID and service location information, we set the maximum number of services to 5,825. This means that each service ID or service location information is mapped from the 5,825 dimension to a $2^E$-dimensional vector through the embedding layer. The user's data is similarly processed.

\autoref{fig:E} shows the experimental results. In dataset D1, the prediction performance becomes more accurate as E increases, and the prediction performance stabilizes when E = 10. However, in D2, with more sparse training data, the prediction performance worsens when E is too large.

\noindent
\textbf{Results:} The default embedding size in this paper is E = 10, as it achieves the best prediction performance in dataset D1.

\section{CONCLUSION}\label{sec:CONCLUSION}
This paper tackles the challenge of QoS prediction errors caused by noisy features data. To overcome this challenge, we propose the use of latent probability space learning networks, or PDS-Net. PDS-Net initializes three different prior distributions through a prior net.
And approximates the posterior distribution from true labels through a condition-based joint loss function.  In general, we first propose targeted solutions to the problem of feature noise in quality prediction of distributed Web systems. Experimental results show that PDS-Net consistently outperforms state-of-the-art methods, with the best predictive performance on both real Qos datasets. In the future, we aim to develop more efficient prediction algorithms that can achieve higher prediction accuracy and handle large-scale datasets.

\ifCLASSOPTIONcaptionsoff
  \newpage
\fi

\bibliographystyle{IEEEtran}
\bibliography{IEEEabrv,my}

\begin{thebibliography}{10}
\providecommand{\url}[1]{#1}
\csname url@samestyle\endcsname
\providecommand{\newblock}{\relax}
\providecommand{\bibinfo}[2]{#2}
\providecommand{\BIBentrySTDinterwordspacing}{\spaceskip=0pt\relax}
\providecommand{\BIBentryALTinterwordstretchfactor}{4}
\providecommand{\BIBentryALTinterwordspacing}{\spaceskip=\fontdimen2\font plus
\BIBentryALTinterwordstretchfactor\fontdimen3\font minus
  \fontdimen4\font\relax}
\providecommand{\BIBforeignlanguage}[2]{{%
\expandafter\ifx\csname l@#1\endcsname\relax
\typeout{** WARNING: IEEEtran.bst: No hyphenation pattern has been}%
\typeout{** loaded for the language `#1'. Using the pattern for}%
\typeout{** the default language instead.}%
\else
\language=\csname l@#1\endcsname
\fi
#2}}
\providecommand{\BIBdecl}{\relax}
\BIBdecl

\bibitem{mouli2016web}
V.~R. Mouli and K.~Jevitha, ``Web services attacks and security-a systematic
  literature review,'' \emph{Procedia Computer Science}, vol.~93, pp. 870--877,
  2016.

\bibitem{menasce2002qos}
D.~A. Menasce, ``Qos issues in web services,'' \emph{IEEE internet computing},
  vol.~6, no.~6, pp. 72--75, 2002.

\bibitem{HeNCF}
X.~He, L.~Liao, H.~Zhang, L.~Nie, X.~Hu, and T.-S. Chua, ``Neural collaborative
  filtering,'' in \emph{Proceedings of the 26th International Conference on
  World Wide Web}, ser. WWW '17.\hskip 1em plus 0.5em minus 0.4em\relax
  Republic and Canton of Geneva, CHE: International World Wide Web Conferences
  Steering Committee, 2017, p. 173–182.

\bibitem{huang2016deep}
G.~Huang, Y.~Sun, Z.~Liu, D.~Sedra, and K.~Q. Weinberger, ``Deep networks with
  stochastic depth,'' in \emph{European conference on computer vision}.\hskip
  1em plus 0.5em minus 0.4em\relax Springer, 2016, pp. 646--661.

\bibitem{carlkadie1998empirical}
J.~B.~D. CarlKadie, ``Empirical analysis of predictive algorithms for
  collaborative filtering,'' \emph{Microsoft Research Microsoft Corporation One
  Microsoft Way Redmond, WA}, vol. 98052, 1998.

\bibitem{13}
B.~Sarwar, G.~Karypis, J.~Konstan, and J.~Riedl, ``Item-based collaborative
  filtering recommendation algorithms,'' in \emph{Proceedings of the 10th
  international conference on World Wide Web}, 2001, pp. 285--295.

\bibitem{18}
M.~Tang, Y.~Jiang, J.~Liu, and X.~Liu, ``Location-aware collaborative filtering
  for qos-based service recommendation,'' in \emph{2012 IEEE 19th international
  conference on web services}.\hskip 1em plus 0.5em minus 0.4em\relax IEEE,
  2012, pp. 202--209.

\bibitem{26}
Y.~Koren, R.~Bell, and C.~Volinsky, ``Matrix factorization techniques for
  recommender systems,'' \emph{Computer}, vol.~42, no.~8, pp. 30--37, 2009.

\bibitem{86}
Z.~Wang, X.~Zhang, M.~Yan, L.~Xu, and D.~Yang, ``Hsa-net: Hidden-state-aware
  networks for high-precision qos prediction,'' \emph{IEEE Transactions on
  Parallel and Distributed Systems}, vol.~33, no.~6, pp. 1421--1435, 2021.

\bibitem{21}
Y.~Zhang, C.~Yin, Q.~Wu, Q.~He, and H.~Zhu, ``Location-aware deep collaborative
  filtering for service recommendation,'' \emph{IEEE Transactions on Systems,
  Man, and Cybernetics: Systems}, 2019.

\bibitem{zou2022ncrl}
G.~Zou, S.~Wu, S.~Hu, C.~Cao, Y.~Gan, B.~Zhang, and Y.~Chen, ``Ncrl:
  Neighborhood-based collaborative residual learning for adaptive qos
  prediction,'' \emph{IEEE Transactions on Services Computing}, 2022.

\bibitem{zhang2021probability}
W.~Zhang, L.~Xu, M.~Yan, Z.~Wang, and C.~Fu, ``A probability distribution and
  location-aware resnet approach for qos prediction,'' \emph{Journal of Web
  Engineering}, vol.~20, no.~4, pp. 1251--1290, 2021.

\bibitem{xie2019feature}
C.~Xie, Y.~Wu, L.~v.~d. Maaten, A.~L. Yuille, and K.~He, ``Feature denoising
  for improving adversarial robustness,'' in \emph{Proceedings of the IEEE/CVF
  conference on computer vision and pattern recognition}, 2019, pp. 501--509.

\bibitem{ye2021outlier}
F.~Ye, Z.~Lin, C.~Chen, Z.~Zheng, and H.~Huang, ``Outlier-resilient web service
  qos prediction,'' in \emph{Proceedings of the Web Conference 2021}, 2021, pp.
  3099--3110.

\bibitem{lu2023feature}
T.~Lu, X.~Zhang, Z.~Wang, and M.~Yan, ``A feature distribution smoothing
  network based on gaussian distribution for qos prediction,'' in \emph{2023
  IEEE International Conference on Web Services (ICWS)}.\hskip 1em plus 0.5em
  minus 0.4em\relax IEEE Computer Society, 2023, pp. 687--694.

\bibitem{miliauskaite2023effect}
J.~Miliauskait{\.e} and D.~Kalibatien{\.e}, ``An effect of user experience on a
  data-driven fuzzy inference of web service quality,'' \emph{INTERNATIONAL
  JOURNAL OF COMPUTERS COMMUNICATIONS \& CONTROL}, vol.~18, no.~4, 2023.

\bibitem{zheng2008ws}
Z.~Zheng and M.~R. Lyu, ``Ws-dream: A distributed reliability assessment
  mechanism for web services,'' in \emph{2008 IEEE International Conference on
  Dependable Systems and Networks With FTCS and DCC (DSN)}.\hskip 1em plus
  0.5em minus 0.4em\relax IEEE, 2008, pp. 392--397.

\bibitem{3}
Z.~Zheng, H.~Ma, M.~R. Lyu, and I.~King, ``Qos-aware web service recommendation
  by collaborative filtering,'' \emph{IEEE Transactions on services computing},
  vol.~4, no.~2, pp. 140--152, 2010.

\bibitem{zhang2023deep}
P.~Zhang, J.~Ren, W.~Huang, Y.~Chen, Q.~Zhao, and H.~Zhu, ``A deep-learning
  model for service qos prediction based on feature mapping and inference,''
  \emph{IEEE Transactions on Services Computing}, 2023.

\bibitem{mohebali2020probabilistic}
B.~Mohebali, A.~Tahmassebi, A.~Meyer-Baese, and A.~H. Gandomi, ``Probabilistic
  neural networks: a brief overview of theory, implementation, and
  application,'' \emph{Handbook of Probabilistic Models}, pp. 347--367, 2020.

\bibitem{kingma2013auto}
D.~P. Kingma and M.~Welling, ``Auto-encoding variational bayes,'' \emph{arXiv
  preprint arXiv:1312.6114}, 2013.

\bibitem{lee2015deeply}
C.-Y. Lee, S.~Xie, P.~Gallagher, Z.~Zhang, and Z.~Tu, ``Deeply-supervised
  nets,'' in \emph{Artificial intelligence and statistics}.\hskip 1em plus
  0.5em minus 0.4em\relax PMLR, 2015, pp. 562--570.

\bibitem{wang2015training}
L.~Wang, C.-Y. Lee, Z.~Tu, and S.~Lazebnik, ``Training deeper convolutional
  networks with deep supervision,'' \emph{arXiv}.

\bibitem{li2020dynamic}
D.~Li and Q.~Chen, ``Dynamic hierarchical mimicking towards consistent
  optimization objectives,'' in \emph{Proceedings of the IEEE/CVF Conference on
  Computer Vision and Pattern Recognition}, 2020, pp. 7642--7651.

\bibitem{zhang2022contrastive}
L.~Zhang, X.~Chen, J.~Zhang, R.~Dong, and K.~Ma, ``Contrastive deep
  supervision,'' in \emph{European Conference on Computer Vision}.\hskip 1em
  plus 0.5em minus 0.4em\relax Springer, 2022, pp. 1--19.

\bibitem{zhang2018deep}
Y.~Zhang and A.~Chung, ``Deep supervision with additional labels for retinal
  vessel segmentation task,'' in \emph{International conference on medical
  image computing and computer-assisted intervention}.\hskip 1em plus 0.5em
  minus 0.4em\relax Springer, 2018, pp. 83--91.

\bibitem{zhang2019scan}
L.~Zhang, Z.~Tan, J.~Song, J.~Chen, C.~Bao, and K.~Ma, ``Scan: A scalable
  neural networks framework towards compact and efficient models,''
  \emph{Advances in Neural Information Processing Systems}, vol.~32, 2019.

\bibitem{zhang2020task}
L.~Zhang, Y.~Shi, Z.~Shi, K.~Ma, and C.~Bao, ``Task-oriented feature
  distillation,'' \emph{Advances in Neural Information Processing Systems},
  vol.~33, pp. 14\,759--14\,771, 2020.

\bibitem{hussain2022new}
W.~Hussain, J.~M. Merig{\'o}, M.~R. Raza, and H.~Gao, ``A new qos prediction
  model using hybrid iowa-anfis with fuzzy c-means, subtractive clustering and
  grid partitioning,'' \emph{Information Sciences}, vol. 584, pp. 280--300,
  2022.

\bibitem{muslim2022s}
H.~S.~M. Muslim, S.~Rubab, M.~M. Khan, N.~Iltaf, A.~K. Bashir, and K.~Javed,
  ``S-rap: relevance-aware qos prediction in web-services and user contexts,''
  \emph{Knowledge and Information Systems}, vol.~64, no.~7, pp. 1997--2022,
  2022.

\bibitem{wu2020data}
D.~Wu, X.~Luo, M.~Shang, Y.~He, G.~Wang, and X.~Wu, ``A
  data-characteristic-aware latent factor model for web services qos
  prediction,'' \emph{IEEE Transactions on Knowledge and Data Engineering},
  vol.~34, no.~6, pp. 2525--2538, 2020.

\bibitem{chowdhury2020cahphf}
R.~R. Chowdhury, S.~Chattopadhyay, and C.~Adak, ``Cahphf: context-aware
  hierarchical qos prediction with hybrid filtering,'' \emph{IEEE Transactions
  on Services Computing}, vol.~15, no.~4, pp. 2232--2247, 2020.

\bibitem{liu2019context}
Z.~Liu, Q.~Z. Sheng, X.~Xu, D.~Chu, and W.~E. Zhang, ``Context-aware and
  adaptive qos prediction for mobile edge computing services,'' \emph{IEEE
  Transactions on Services Computing}, vol.~15, no.~1, pp. 400--413, 2019.

\bibitem{zheng2020web}
Z.~Zheng, X.~Li, M.~Tang, F.~Xie, and M.~R. Lyu, ``Web service qos prediction
  via collaborative filtering: A survey,'' \emph{IEEE Transactions on Services
  Computing}, vol.~15, no.~4, pp. 2455--2472, 2020.

\bibitem{liang2021recurrent}
T.~Liang, M.~Chen, Y.~Yin, L.~Zhou, and H.~Ying, ``Recurrent neural network
  based collaborative filtering for qos prediction in iov,'' \emph{IEEE
  Transactions on Intelligent Transportation Systems}, vol.~23, no.~3, pp.
  2400--2410, 2021.

\bibitem{li2021topology}
J.~Li, H.~Wu, J.~Chen, Q.~He, and C.-H. Hsu, ``Topology-aware neural model for
  highly accurate qos prediction,'' \emph{IEEE Transactions on Parallel and
  Distributed Systems}, vol.~33, no.~7, pp. 1538--1552, 2021.

\bibitem{xia2021joint}
Y.~Xia, D.~Ding, Z.~Chang, and F.~Li, ``Joint deep networks based multi-source
  feature learning for qos prediction,'' \emph{IEEE Transactions on Services
  Computing}, vol.~15, no.~4, pp. 2314--2327, 2021.

\bibitem{ghafouri2020survey}
S.~H. Ghafouri, S.~M. Hashemi, and P.~C. Hung, ``A survey on web service qos
  prediction methods,'' \emph{IEEE Transactions on Services Computing},
  vol.~15, no.~4, pp. 2439--2454, 2020.

\bibitem{1}
K.~Lee, J.~Park, and J.~Baik, ``Location-based web service qos prediction via
  preference propagation for improving cold start problem,'' in \emph{2015 IEEE
  International Conference on Web Services}.\hskip 1em plus 0.5em minus
  0.4em\relax IEEE, 2015, pp. 177--184.

\bibitem{2}
X.~Chen, X.~Liu, Z.~Huang, and H.~Sun, ``Regionknn: A scalable hybrid
  collaborative filtering algorithm for personalized web service
  recommendation,'' in \emph{2010 IEEE international conference on web
  services}.\hskip 1em plus 0.5em minus 0.4em\relax IEEE, 2010, pp. 9--16.

\bibitem{4}
Z.~Chen, L.~Shen, F.~Li, and D.~You, ``Your neighbors alleviate cold-start: On
  geographical neighborhood influence to collaborative web service qos
  prediction,'' \emph{Knowledge-Based Systems}, vol. 138, pp. 188--201, 2017.

\bibitem{12}
Z.~Tan and L.~He, ``An efficient similarity measure for user-based
  collaborative filtering recommender systems inspired by the physical
  resonance principle,'' \emph{IEEE Access}, vol.~5, pp. 27\,211--27\,228,
  2017.

\bibitem{14}
H.~Ma, I.~King, and M.~R. Lyu, ``Effective missing data prediction for
  collaborative filtering,'' in \emph{Proceedings of the 30th annual
  international ACM SIGIR conference on Research and development in information
  retrieval}, 2007, pp. 39--46.

\bibitem{79}
J.~A. Konstan, B.~N. Miller, D.~Maltz, J.~L. Herlocker, L.~R. Gordon, and
  J.~Riedl, ``Grouplens: Applying collaborative filtering to usenet news,''
  \emph{Communications of the ACM}, vol.~40, no.~3, pp. 77--87, 1997.

\bibitem{8}
X.~Luo, M.~Zhou, S.~Li, Z.~You, Y.~Xia, and Q.~Zhu, ``A nonnegative latent
  factor model for large-scale sparse matrices in recommender systems via
  alternating direction method,'' \emph{IEEE transactions on neural networks
  and learning systems}, vol.~27, no.~3, pp. 579--592, 2015.

\bibitem{9}
Y.~Shi, M.~Larson, and A.~Hanjalic, ``Collaborative filtering beyond the
  user-item matrix: A survey of the state of the art and future challenges,''
  \emph{ACM Computing Surveys (CSUR)}, vol.~47, no.~1, pp. 1--45, 2014.

\bibitem{35}
Y.~Zhang, Z.~Zheng, and M.~R. Lyu, ``Wspred: A time-aware personalized qos
  prediction framework for web services,'' in \emph{2011 IEEE 22nd
  International Symposium on Software Reliability Engineering}.\hskip 1em plus
  0.5em minus 0.4em\relax IEEE, 2011, pp. 210--219.

\bibitem{81}
S.~Wang, Y.~Zhao, L.~Huang, J.~Xu, and C.-H. Hsu, ``Qos prediction for service
  recommendations in mobile edge computing,'' \emph{Journal of Parallel and
  Distributed Computing}, vol. 127, pp. 134--144, 2019.

\bibitem{29}
Z.~Luo, L.~Liu, J.~Yin, Y.~Li, and Z.~Wu, ``Latent ability model: A generative
  probabilistic learning framework for workforce analytics,'' \emph{IEEE
  Transactions on Knowledge and Data Engineering}, vol.~31, no.~5, pp.
  923--937, 2018.

\bibitem{20}
X.~He, L.~Liao, H.~Zhang, L.~Nie, X.~Hu, and T.-S. Chua, ``Neural collaborative
  filtering,'' in \emph{Proceedings of the 26th international conference on
  world wide web}, 2017, pp. 173--182.

\bibitem{83}
Q.~Zhou, H.~Wu, K.~Yue, and C.-H. Hsu, ``Spatio-temporal context-aware
  collaborative qos prediction,'' \emph{Future Generation Computer Systems},
  vol. 100, pp. 46--57, 2019.

\bibitem{wang2016online}
H.~Wang, L.~Wang, Q.~Yu, Z.~Zheng, A.~Bouguettaya, and M.~R. Lyu, ``Online
  reliability prediction via motifs-based dynamic bayesian networks for
  service-oriented systems,'' \emph{IEEE Transactions on Software Engineering},
  vol.~43, no.~6, pp. 556--579, 2016.

\bibitem{xiong2018deep}
R.~Xiong, J.~Wang, N.~Zhang, and Y.~Ma, ``Deep hybrid collaborative filtering
  for web service recommendation,'' \emph{Expert systems with Applications},
  vol. 110, pp. 191--205, 2018.

\bibitem{im2017denoising}
D.~Im~Im, S.~Ahn, R.~Memisevic, and Y.~Bengio, ``Denoising criterion for
  variational auto-encoding framework,'' in \emph{Proceedings of the AAAI
  conference on artificial intelligence}, vol.~31, no.~1, 2017.

\bibitem{77}
F.~Ye, Z.~Lin, C.~Chen, Z.~Zheng, and H.~Huang, ``Outlier-resilient web service
  qos prediction,'' in \emph{Proceedings of the Web Conference 2021}, 2021, pp.
  3099--3110.

\bibitem{41}
J.~M. Hern{\'a}ndez-Lobato, N.~Houlsby, and Z.~Ghahramani, ``Probabilistic
  matrix factorization with non-random missing data,'' in \emph{International
  Conference on Machine Learning}, 2014, pp. 1512--1520.

\bibitem{40}
Y.~Yin, L.~Chen, Y.~Xu, J.~Wan, H.~Zhang, and Z.~Mai, ``Qos prediction for
  service recommendation with deep feature learning in edge computing
  environment,'' \emph{Mobile Networks and Applications}, pp. 1--11, 2019.

\bibitem{wu2022double}
D.~Wu, P.~Zhang, Y.~He, and X.~Luo, ``A double-space and double-norm ensembled
  latent factor model for highly accurate web service qos prediction,''
  \emph{IEEE Transactions on Services Computing}, vol.~16, no.~2, pp. 802--814,
  2022.

\bibitem{zou2024frln}
G.~Zou, W.~Yu, S.~Hu, Y.~Gan, B.~Zhang, and Y.~Chen, ``Frln: Federated residual
  ladder network for data-protected qos prediction,'' \emph{IEEE Transactions
  on Services Computing}, 2024.

\end{thebibliography}
\vspace{-1cm}

\vspace{-11cm}
\end{document}